\documentclass[twocolumn]{aastex61}

\shorttitle{$\mathrm{C^{18}O}$ Clumps in Cygnus~X}
\shortauthors{Takekoshi et al.}

\begin{document}

\title{Nobeyama 45-m Cygnus X CO Survey: II Physical Properties of $\mathrm{C^{18}O}$ Clumps}

\correspondingauthor{Tatsuya Takekoshi}
\author[0000-0002-4124-797X]{Tatsuya Takekoshi}
\affiliation{Institute of Astronomy, The University of Tokyo, 2-21-1 Osawa, Mitaka, Tokyo 181-0015, Japan}
\affiliation{Graduate School of Informatics and Engineering, The University of Electro-Communications, 1-5-1 Chofugaoka, Chofu,Tokyo 182-8585, Japan}
\email{tatsuya.takekoshi@ioa.s.u-tokyo.ac.jp}
\author[0000-0002-6375-7065]{Shinji Fujita}
\affiliation{Department of Astrophysics, Nagoya University, Chikusa-ku, Nagoya 464-8602, Japan}
\author[0000-0003-0732-2937]{Atsushi Nishimura}
\affiliation{Department of Physical Science, Osaka Prefecture University, Gakuen 1-1, Sakai, Osaka 599-8531, Japan}
\author[0000-0003-4402-6475]{Kotomi Taniguchi}
\affiliation{Nobeyama Radio Observatory, National Astronomical Observatory of Japan (NAOJ), National Institutes of Natural Sciences (NINS), 462-2, Nobeyama, Minamimaki, Minamisaku, Nagano 384-1305, Japan}
\affiliation{Departments of Astronomy and Chemistry, University of Virginia, Charlottesville, VA 22904, USA}
\author{Mitsuyoshi Yamagishi}
\affiliation{Institute of Space and Astronautical Science, Japan Aerospace Exploration Agency, Chuo-ku, Sagamihara 252-5210, Japan}
\author{Mitsuhiro Matsuo}
\affiliation{Nobeyama Radio Observatory, National Astronomical Observatory of Japan (NAOJ), National Institutes of Natural Sciences (NINS), 462-2, Nobeyama, Minamimaki, Minamisaku, Nagano 384-1305, Japan}
\author{Satoshi Ohashi}
\affiliation{The Institute of Physical and Chemical Research (RIKEN), 2-1, Hirosawa, Wako-shi, Saitama 351-0198, Japan}
\author[0000-0002-2062-1600]{Kazuki Tokuda}
\affiliation{Department of Physical Science, Osaka Prefecture University, Gakuen 1-1, Sakai, Osaka 599-8531, Japan}
\affiliation{National Astronomical Observatory of Japan (NAOJ), National Institutes of Natural Sciences (NINS), 2-21-1, Osawa, Mitaka, Tokyo 181-8588, Japan}
\author[0000-0001-9778-6692]{Tetsuhiro Minamidani}
\affiliation{Nobeyama Radio Observatory, National Astronomical Observatory of Japan (NAOJ), National Institutes of Natural Sciences (NINS), 462-2, Nobeyama, Minamimaki, Minamisaku, Nagano 384-1305, Japan}
\affiliation{Department of Astronomical Science, School of Physical Science, SOKENDAI (The Graduate University for Advanced Studies), 2-21-1, Osawa, Mitaka, Tokyo 181-8588, Japan}



\begin{abstract}
We report the statistical physical properties of the C$^{18}$O($J=1-0$) clumps present in a prominent cluster-forming region, Cygnus~X, using the dataset obtained by the Nobeyama 45-m radio telescope.
This survey covers 9 deg$^2$ of the north and south regions of Cygnus~X, and totally 174 C$^{18}$O clumps are identified using the dendrogram method.
Assuming a distance of 1.4~kpc, these clumps have radii of 0.2--1~pc, velocity dispersions of $<2.2~\mathrm{km~s^{-1}}$, gas masses of 30--3000~$M_\sun$, and H$_2$ densities of (0.2--5.5)$\times10^4~\mathrm{cm^{-3}}$. 
We confirm that the C$^{18}$O clumps in the north region have a higher H$_2$ density than those in the south region, supporting the existence of a difference in the evolution stages, consistent with the star formation activity of these regions.
The difference in the clump properties of the star-forming and starless clumps is also confirmed by the radius, velocity dispersion, gas mass, and H$_2$ density. 
The average virial ratio of 0.3 supports that these clumps are gravitationally bound.
The C$^{18}$O clump mass function shows two spectral index components, $\alpha=-1.4$ in 55--140~$M_\sun$ and $\alpha=-2.1$ in $>140~M_\sun$, which are consistent with the low- and intermediate-mass parts of the Kroupa's initial mass function.
The spectral index in the star-forming clumps in $>140~M_\sun$ is consistent with that of the starless clumps in 55--140~$M_\sun$, suggesting that the latter will evolve into star-forming clumps while retaining the gas accretion.
Assuming a typical star formation efficiency of molecular clumps (10\%), about ten C$^{18}$O clumps having a gas mass of $>10^3~M_\sun$ will evolve into open clusters containing one or more OB stars.
\end{abstract}


\section{Introduction}
\label{sec:intro}
Stars are formed in dense molecular cores and clumps, which are defined as compact ($\sim$0.1, and 1~pc, respectively) and dense ($\ga 10^4$--$10^5~\mathrm{H_2~cm^{-3}}$) structures \citep{2018ARA&A..56...41M,2016ApJ...833..209O,2009ApJ...696..268Z,2000prpl.conf...97W}.
Understanding the physical and chemical properties of dense cores and clumps is one of the most important astrophysical topics in regard to the star formation process connecting a molecular cloud and protostar, mechanism to determine the initial stellar mass function (IMF), and enrichment of interstellar molecules.
Therefore, observational studies of dense cores toward various star-forming regions have been conducted by some methods.
In particular, thermal dust continuum observations using space and ground-based imaging arrays at sub-millimeter wavelengths identify hundreds of dense cores in nearby star-forming regions and reveal their statistical properties \citep[e.g.,][]{1998A&A...336..150M,2006ApJ...638..293E}.
Dust extinction also provides a similar observational approach to exhibit the statistical properties of the dense cores in nearby molecular clouds\citep[e.g.,][]{2007A&A...462L..17A}. 

Concurrently, molecular line mapping using dense molecular gas tracers, such as C$^{18}$O, CS, NH$_3$, and H$^{13}$CO$^+$, is a complementary approach to reveal the dense core and clump properties of star-forming regions. This method assists in decomposing spatially overlapping components using the velocity information and in diagnosing the kinematic properties of the dense cores \citep[e.g.,][]{1983ApJ...266..309M,1983ApJ...264..517M,1990ApJ...356..513S,1993ApJ...404..643T,1996ApJ...465..815O,2002ApJ...575..950O,2004A&A...416..191T}.
In particular, C$^{18}$O ($J=1$--$0$) surveys performed by the 4-m NANTEN and Nagoya University telescopes were promoted toward the high-density regions traced by a $^{13}$CO line \citep[e.g.][]{1996ApJ...465..815O,1999PASJ...51..895H,2000ApJ...528..817T,2002ApJ...575..950O} for nearby low-mass star-forming regions ($<200$~pc).
The C$^{18}$O cores identified by these observations exhibit densities of $10^4$--$10^5~\mathrm{cm^{-3}}$, radii of 0.1--0.5~pc, and gas masses of 1--100~$M_\odot$, typically with a spatial resolution of $\sim 0.1$~pc.

By contrast, understanding of the statistical core and clump properties in high-mass star-forming regions using molecular lines is still limited.
Although C$^{18}$O mapping observations have investigated the statistical core and clump properties in some active centers of high-mass star-forming regions \citep[e.g.,][]{1990ApJ...356..513S,1998A&A...329..249K,2008MNRAS.386.1069W,2009ApJ...705L..95I,2011ApJ...732..101I}, the observed field size in each case is only a few hundred square arcminutes.
Thus, previous studies are probably biased to the core and clump properties in massive filaments. Such filaments are expected to be affected by the efficient mass accretion by the large-scale conversing flows \citep{2016A&A...592A..54A,2018arXiv181100812F,2018arXiv181104400T} as well as by the strong feedback and phenomena associated with high-mass star formation, such as strong ultra-violet (UV) radiation feedback, jet and outflow from a protostar, and supernova explosions.
Therefore, determining the physical properties of the C$^{18}$O cores and clumps in high-mass star-forming regions with a large-field ($>1~\mathrm{deg}^2$) survey is important to obtain the complete mechanism of a high-mass star formation in giant molecular clouds (GMCs).

Cygnus~X is one of the most massive complexes of GMCs in our galaxy.
Its environment is characterized by the current extremely active star formation, and a close distance of 1.4~kpc\citep{2012A&A...539A..79R} provides an opportunity to investigate the cluster formation process and ISM affected by strong stellar UV feedback by high-mass stars.
At the center of the Cygnus~X complex is the Cygnus OB2 association, which is known as one of the most massive associations of young stars in our galaxy containing $>200$ OB stars \citep{2015MNRAS.449..741W}.
The total molecular gas mass of Cygnus~X is estimated to be $3\times 10^6~\mathrm{M_\odot}$\citep{2006A&A...458..855S}.

Cygnus~X is divided into northern and southern molecular cloud complexes (hereafter referred as North and South, respectively), which have $^{13}$CO($J=2-1$)-traced dense gas masses of 2 and 3$\times 10^5\ \mathrm{M_\sun}$, respectively \citep{2006A&A...458..855S}.
Cygnus~X North shows an extremely filamentary structure of dust and molecular gas \citep[e.g.,][]{2011A&A...529A...1S, 2016A&A...591A..40S}, and contains well-known star-forming regions represented as DR21 and W75N, consisting of numerous fragmentary structures and massive dense cores, which can form high-mass stars.
Compared to Cygnus~X North, the South region shows a relatively weak star-forming activity, but the existence of a large amount of molecular gas component suggests the possibility of formation of stellar clusters in the South region \citep{2018ApJS..235....9Y,2006A&A...458..855S}. 
Thus, the Cygnus~X complex, which contains these various environments, is the best target to investigate the star formation process from molecular gases to a massive stellar cluster and the stellar radiation feedback to the GMCs.
Previously, molecular gas distribution in Cygnus~X was investigated by $^{12/13}$CO lines \citep{2006A&A...458..855S,2011A&A...529A...1S}, and \textit{Herschel} revealed the dust distribution via imaging of the sub-millimeter continuum emission\citep{2016A&A...591A..40S}.
However, it is important to investigate the gas properties of dense clumps using an optically thin line emission to reveal the formation process of high-mass stars and stellar clusters from molecular cores.

In this study, we investigated C$^{18}$O clump properties using multi-line CO and CN survey data at 3~mm wavelength toward the main part of the Cygnus~X GMC complex using the Nobeyama 45~m telescope \citep{2018ApJS..235....9Y}.
Section~\ref{sec2} describes the details of the C$^{18}$O observation, clump identification method, and estimation of the physical properties of the identified clumps.
The result of data analysis is presented in Section~\ref{sec3}.
In Section~\ref{sec4}, we discuss the physical properties of the C$^{18}$O clumps obtained by this survey.
Finally, Section~\ref{sec5} summarizes the main results of this study.

\section{Observation and Analysis}
\label{sec2}
\subsection{FOREST/NRO 45-m Data}
The C$^{18}$O ($J=$1--0) data of the Cygnus~X region were obtained by the FOREST receiver \citep{minamidani2016development} mounted on the Nobeyama 45~m radio telescope, along with the $^{12}$CO ($J=$1--0), $^{13}$CO ($J=$1--0), and CN ($N=$1--0) data\footnote{The datasets are publicly available at \url{https://cygnus45.github.io}} \citep{2018ApJS..235....9Y}.
The observations covered a $\sim9$~deg$^2$ field, which included the main parts of the North and South GMCs, by connecting $1^\circ \times 1^\circ$ patches \citep[FUGIN scan,][]{2017PASJ...69...78U}.
The angular resolution of the telescope was $\sim 16''$ FWHM at the C$^{18}$O band.

To improve the sensitivity, we convolved the cube to a spatial resolution of $46''$ FWHM with a pixel grid of $22.7''$ and binned to a velocity resolution of $0.25~\mathrm{km~s^{-1}}$.
Consequently, the median rms noise level of the final C$^{18}$O image $T_\mathrm{rms}$ became 0.35~K on the $T_\mathrm{mb}$ scale.
The observation and data analysis procedures are described in \citet{2018ApJS..235....9Y} in detail.

\subsection{C$^{18}$O clump identification}
\label{sec2.2}
We identify C$^{18}$O clumps from the C$^{18}$O cube using the \texttt{astrodendro} package, which is based on the dendrogram algorithm \citep{2008ApJ...679.1338R}.
A dendogram is used to construct a tree structure consisting of trunks, branches, and leaves.
A ``trunk'' is defined as a set of voxels such that $T_\mathrm{mb}$ is larger than $T_\mathrm{min}$ and the voxel number, $n_\mathrm{vox}$, is not less than the integer, $n_\mathrm{vox}^\mathrm{min}$.
A trunk is split into one or more leaves by a ``branch'', which is a node of more compact structures (leaf or branch).
A ``leaf'' is defined as a local peak such that its height is higher than $T_\mathrm{delta}$ from the skirt of the peak and the voxel number is not less than $n_\mathrm{vox}^\mathrm{min}$.
From the definition, leaves are identified as compact clumps that do not have multiple peaks, and therefore, a dendrogram is available as an identification algorithm for molecular clumps and cores.
\citet{2018ApJ...853..160C} used the dendrogram method to identify the dense cores and clumps in G~286.21+0.27 and reported that the dendrogram-identified cores showed a spectral index of the core mass function that was more consistent with the Salpeter-IMF than that of the clumpfind-identified cores.
Thus, it is reasonable to adopt the dendrogram algorithm as a core and clump identification procedure.

In the analysis, we adopted $T_\mathrm{min}=3T_\mathrm{rms}$ and $T_\mathrm{delta}=2T_\mathrm{rms}$ to identify C$^{18}$O clumps with a reliable signal-to-noise ratio.
We also used $n_\mathrm{vox}^\mathrm{min}=16$ to avoid false detection of clumps by picking up random noise.

\subsection{Physical property estimation}
\label{sec2.3}
We estimated C$^{18}$O clump properties such as radius $R_\mathrm{cl}$, local thermal equilibrium (LTE) mass $M_\mathrm{LTE}$, FWHM velocity width $dv_\mathrm{cl}$, and virial mass $M_\mathrm{vir}$.
We followed the method generally adopted by previous C$^{18}$O core studies \citep{1996ApJ...465..815O,2009ApJ...705L..95I, 2011ApJ...732..101I, 2015ApJS..217....7S} and made further refinements to improve the reliability of the estimated physical properties, motivated by \citet{2015ApJS..216...18N}.

The radius of a C$^{18}$O clump is defined using the pixel number projected on the sky ($l$-$b$ plane of the galactic coordinate), $n_\mathrm{sky}$, assuming a spherically symmetric clump: 
\begin{equation}
R_\mathrm{cl}= D\theta_\mathrm{pix}\sqrt{\frac{n_\mathrm{sky}}{\pi}},
\end{equation}
where $\theta_\mathrm{pix}=22.7''$ is the pixel grid spacing along with the galactic coordinate of the data cube and $D$ is the distance to the clump from the solar system.

We define the integrated intensity of an object as
\begin{equation}
I=\sum^{n_\mathrm{vox}}_{i=1}T_\mathrm{mb}^i dv_\mathrm{spec} (D \theta_\mathrm{pix})^2 
\end{equation}
where $i$ is the voxel number of each core, 
$T_\mathrm{mb}^i$ is the main beam temperature at voxel $i$ and $dv_\mathrm{spec}=0.25~\mathrm{km~s^{-1}}$ is the spectral velocity width. 
The LTE mass is estimated using the equation:
\begin{equation}
M_\mathrm{LTE}(\mathrm{M_\odot})=13.2 \frac{X_\mathrm{C^{18}O}}{5.9\times10^6}\frac{T_\mathrm{ex}}{\mathrm{1~K}} e^{\frac{\mathrm{5.27~K}}{T_\mathrm{ex}}} \frac{I}{\mathrm{1~K~km~s^{-1}~pc^2}},
\end{equation}
where $T_\mathrm{ex}$ is the excitation temperature of the C$^{18}$O molecules. 
We assumed that $^{12}$CO was optically thick and the $T_\mathrm{ex}$ of $^{12}$CO and C$^{18}$O was the same. Thus,
\begin{equation}
T_\mathrm{ex} (\mathrm{K}) =\frac{5.53}{\ln{(1+\frac{5.53}{T_\mathrm{mb}^{\mathrm{^{12}CO}}+0.83})}},
\end{equation}
where $T_\mathrm{mb}^{\mathrm{^{12}CO}} = \max\{T^{\mathrm{^{12}CO},~i}_{\mathrm{mb}}; i=1,2, ..., n_\mathrm{vox}\}$.
We used the isotopic abundance ratio of the C$^{18}$O molecules relative to the $\mathrm{H_2}$ molecules, $X_\mathrm{C^{18}O}=5.9\times 10^6$\citep{1982ApJ...262..590F}. 
Assuming a spherical clump shape, we also defined the mean gas density as
\begin{equation}
n_\mathrm{H_2} = \frac{3M_\mathrm{LTE}}{4\pi 2\mu m_\mathrm{p} {R_\mathrm{cl}}^3},
\end{equation}
where $\mu=1.36$ is the mean molecular weight per proton and $\mathrm{m_p}=1.67\times 10^{-24}~\mathrm{g}$ is a proton mass.

To calculate the virial masses, we removed the effect of the velocity width broadening by the limitation of the spectral resolution, $dv_\mathrm{spec}$. 
The spectral window function obtained by a SAM45 spectrometer can be approximated by a rectangular window function.
Thus, we estimated the actual velocity widths of the clumps, $dv_\mathrm{cl}$, by deconvolving the intensity-weighted velocity dispersions, $dv_\mathrm{obs}$, with a rectangular function.

Following \citet{1987ApJ...319..730S} and \citet{2013ARA&A..51..207B}, we estimated the virial masses of the identified clumps with the radial density profile of $\rho(R)\propto R^{-k}$: 
\begin{equation}
M_\mathrm{vir}=\frac{3(5-2k)}{G(3-k)}R_\mathrm{cl}{\sigma_\mathrm{cl}}^2,
\end{equation}
where $G$ is the gravitational constant, $k$ is a parameter of the density profile, and $\sigma_\mathrm{cl}=dv_\mathrm{cl}/(2\sqrt{2\ln2})$, which is assuming a Gaussian line profile.
We used $k=0$ for the calculation, which assumes a spherically uniform clump that has no external pressure, by following the previous C$^{18}$O studies \citep[e.g.,][]{2009ApJ...705L..95I, 2011ApJ...732..101I, 2015ApJS..217....7S,2002A&A...385..909T}.
The possible bias of the virial mass estimate by the selection of $k$ is a factor of 1--1.4 in the possible range of $0<k<2$ for an isothermal gas sphere \citep[e.g.,][]{1977ApJ...214..488S}.
We also defined the virial ratio, $\alpha_\mathrm{vir}\equiv M_\mathrm{vir}/M_\mathrm{LTE}$.

Finally, we used the canonical distance, $D=1.4$~kpc, from the sun to the clumps, which was determined by parallax measurements by a very long baseline interferometry observation toward four major star-forming regions, as a representative value of the Cygnus~X GMC complex.
The exception was the distance to a corresponding object of background star-forming region AFGL~2592 at $D=3.3$~kpc \citep{2012A&A...539A..79R}.

\newpage
\section{Result}
\label{sec3}
\begin{figure*}[!ht]
\epsscale{1.1}
\plotone{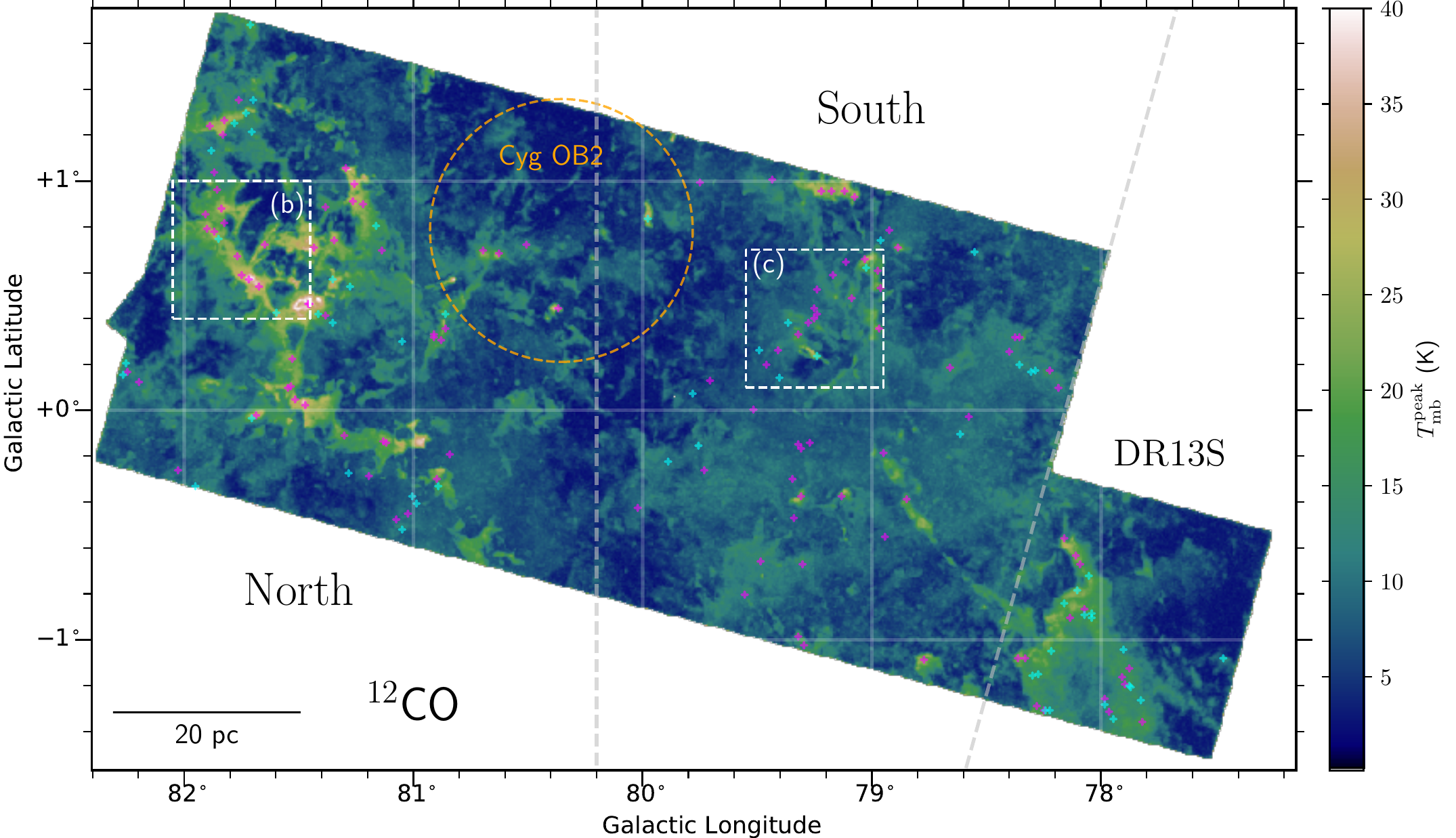}
\caption{The $^{12}$CO peak temperature map obtained by the Nobeyama 45-m Cygnus~X survey. Magenta and cyan crosses show the center positions of the identified protostar-hosted and starless clumps, respectively. An orange dashed-line circle shows the position of the over-dense regions of the OB stars belonging to the Cygnus OB2 association \citep[$\sim$14 pc radius,][]{2015MNRAS.449..741W}. White dashed-line rectangular regions correspond to zoomed images of the (b) DR21/W75N and (c) DR15 regions in Figure \ref{fig:map}.\label{fig:map12CO}}
\end{figure*}

\begin{figure*}[!ht]
\epsscale{1.1}
\plotone{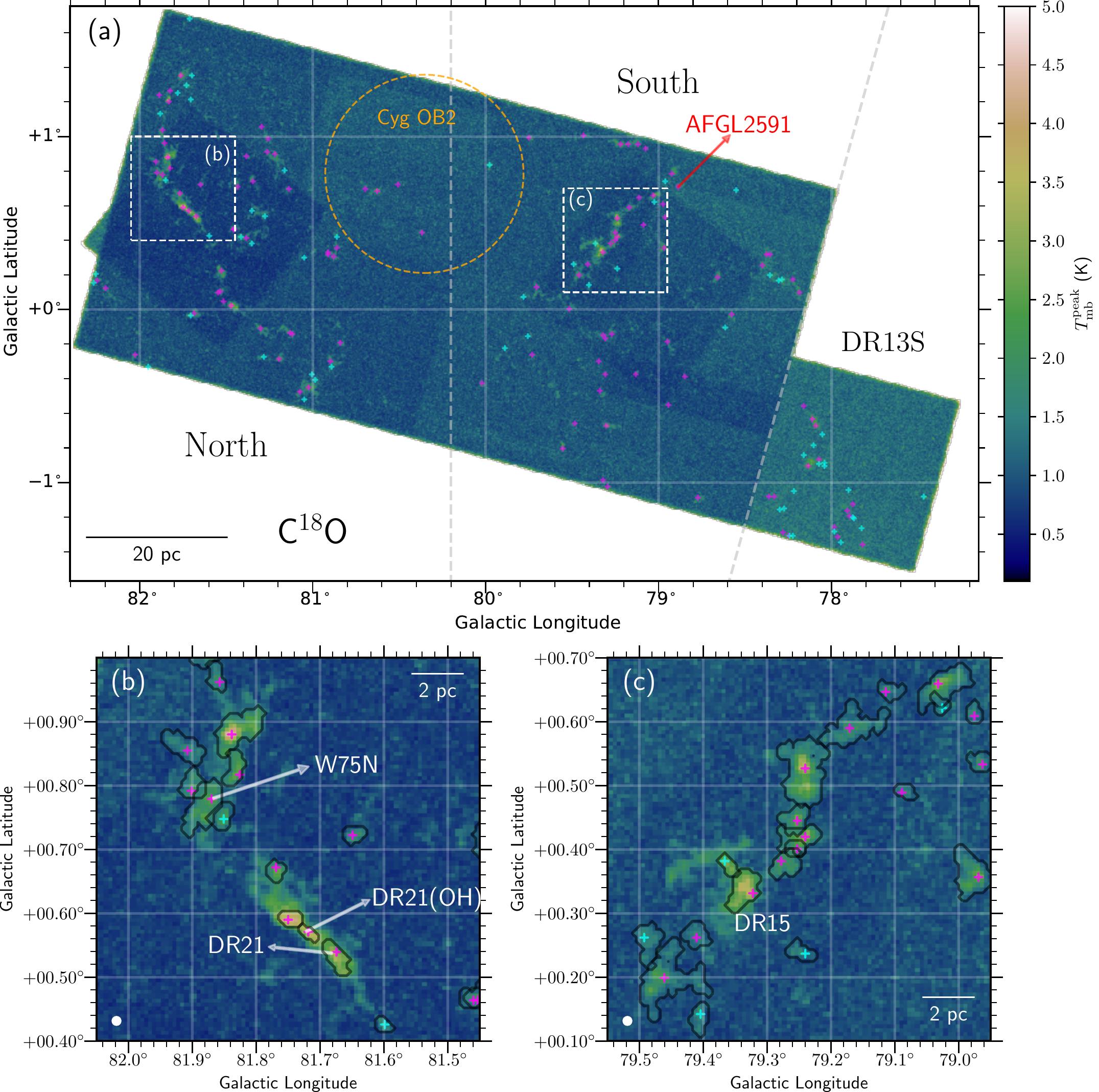}
\caption{The C$^{18}$O peak temperature maps obtained by the Nobeyama 45-m Cygnus~X survey. (a) Overall view of our survey area. The zoomed images of the (b) DR21/W75N and (c) DR15 regions. Black contours show the identified regions as clumps. The other lines and signs are the same as in Figure~\ref{fig:map12CO}. \label{fig:map}}
\end{figure*}

Figures~\ref{fig:map12CO} and \ref{fig:map} show the C$^{12}$O and C$^{18}$O peak main beam temperature images obtained by the observation.
The dense gas traced by the C$^{18}$O emission shows a filamentary and compact distribution.
By contrast, the $^{12}$CO emission traces a more diffuse gas component than the C$^{18}$O emission.

Based on the dendrogram analysis, we identified 177 C$^{18}$O clump candidates as leaves.
From these samples, we excluded three clump candidates that had no corresponding $^{13}$CO emission at the velocity of the C$^{18}$O line, because these would be false detections.
Thus, we identified 174 C$^{18}$O clumps in total.
The positions of the C$^{18}$O clumps are shown in Figures~\ref{fig:map12CO} and \ref{fig:map}.
The C$^{18}$O clump catalog is provided in Table~\ref{table:catalog}.

\begin{longrotatetable}
\begin{deluxetable*}{ccccccccccccccc}
\tablecaption{C$^{18}$O core catalog of the Cygnus-X survey \label{table:catalog}}
\tablewidth{0pt}
\tabletypesize{\scriptsize}
\tablehead{
\colhead{ID} & \colhead{$l_\mathrm{peak}$} & \colhead{$b_\mathrm{peak}$} & \colhead{$v_\mathrm{peak}$} & \colhead{$T_\mathrm{mb}^\mathrm{C^{18}O}$} & \colhead{$T_\mathrm{mb}^\mathrm{^{12}CO}$} & \colhead{$I^\mathrm{C^{18}O}$} & \colhead{$R_\mathrm{cl}$} & \colhead{$M_\mathrm{LTE}$} & \colhead{$n_\mathrm{H_2}$}& \colhead{$dv_\mathrm{cl}$} & \colhead{$M_\mathrm{vir}$} & \colhead{SF?} & \colhead{Region} & \colhead{Edge?}\\
\colhead{} & \colhead{($^\circ$)} & \colhead{($^\circ$)} & \colhead{(km~s$^{-1}$)} & \colhead{(K)} & \colhead{(K)} & \colhead{(K~km~s$^{-1}$~pc$^2$)} & \colhead{(pc)} & \colhead{($10^2~M_\odot$)} & \colhead{($10^4$~cm$^{-3}$)} & \colhead{(km~s$^{-1}$)} & \colhead{($10^2~M_\odot$)} & \colhead{} & \colhead{} & \colhead{}
}
\decimalcolnumbers
\startdata
1 & 80.5080  & 0.7222  & -30.12  & 1.96  $\pm$ 0.39  & 10.07  $\pm$ 1.79  & 0.294  $\pm$ 0.029  & 0.39  & 0.78 & 0.47  & 1.00  & 0.83  & Y & North & \nodata \\
2 & 79.9782  & 0.8357  & -10.12  & 2.52  $\pm$ 0.43  & 32.13  $\pm$ 4.88  & 0.675  $\pm$ 0.068  & 0.43  & 3.69 & 1.70  & 0.91  & 0.76  & \nodata & South & \nodata \\
3 & 78.8873  & 0.7096  & -6.12  & 1.98  $\pm$ 0.47  & 28.24  $\pm$ 4.37  & 7.069  $\pm$ 0.707  & 1.43  & 35.10 & 0.42  & 1.47  & 6.57  & Y & AFGL2592 & \nodata \\
4 & 81.5485  & 0.0980  & -6.38  & 2.00  $\pm$ 0.36  & 24.89  $\pm$ 3.80  & 0.395  $\pm$ 0.040  & 0.36  & 1.79 & 1.38  & 0.76  & 0.45  & Y & North & \nodata \\
5 & 81.4728  & 0.0223  & -4.38  & 3.98  $\pm$ 0.48  & 37.64  $\pm$ 5.69  & 4.619  $\pm$ 0.462  & 0.82  & 28.60 & 1.88  & 1.34  & 3.12  & Y & North & \nodata \\
6 & 81.5170  & 0.0476  & -6.38  & 2.16  $\pm$ 0.34  & 27.83  $\pm$ 4.23  & 0.134  $\pm$ 0.013  & 0.25  & 0.66 & 1.58  & 0.42  & 0.10  & Y & North & \nodata \\
7 & 81.3026  & -0.1101  & -5.88  & 2.42  $\pm$ 0.41  & 22.27  $\pm$ 3.43  & 1.458  $\pm$ 0.146  & 0.81  & 6.10 & 0.41  & 0.65  & 0.75  & Y & North & \nodata \\
8 & 81.1197  & -0.1416  & -5.12  & 2.55  $\pm$ 0.42  & 24.50  $\pm$ 3.76  & 0.532  $\pm$ 0.053  & 0.37  & 2.38 & 1.69  & 0.66  & 0.35  & Y & North & \nodata \\
9 & 81.1323  & -0.1353  & -3.88  & 2.22  $\pm$ 0.40  & 20.15  $\pm$ 3.12  & 0.759  $\pm$ 0.076  & 0.57  & 2.97 & 0.57  & 1.17  & 1.66  & Y & North & \nodata \\
10 & 77.4680  & -1.0811  & -3.88  & 2.56  $\pm$ 0.51  & 14.84  $\pm$ 2.44  & 0.787  $\pm$ 0.079  & 0.45  & 2.54 & 0.98  & 1.18  & 1.33  & \nodata & DR13S & \nodata \\
11 & 81.7503  & 0.5898  & -4.12  & 4.29  $\pm$ 0.53  & 30.25  $\pm$ 4.60  & 1.036  $\pm$ 0.104  & 0.33  & 5.42 & 5.60  & 1.35  & 1.25  & Y & DR13S & \nodata \\
12 & 80.6278  & 0.6844  & -3.38  & 3.10  $\pm$ 0.49  & 25.86  $\pm$ 3.97  & 2.048  $\pm$ 0.205  & 0.71  & 9.53 & 0.94  & 1.96  & 5.74  & Y & North & \nodata \\
13 & 81.5989  & 0.4259  & -3.88  & 1.94  $\pm$ 0.33  & 31.04  $\pm$ 4.70  & 0.155  $\pm$ 0.016  & 0.26  & 0.83 & 1.66  & 0.42  & 0.11  & \nodata & North & \nodata \\
14 & 78.3321  & -1.0811  & -3.12  & 1.92  $\pm$ 0.60  & 24.27  $\pm$ 3.78  & 0.280  $\pm$ 0.028  & 0.40  & 1.24 & 0.70  & 0.75  & 0.49  & Y & South & \nodata \\
15 & 81.8512  & 0.7475  & -3.12  & 2.47  $\pm$ 0.38  & 28.25  $\pm$ 4.29  & 0.614  $\pm$ 0.061  & 0.43  & 3.05 & 1.40  & 0.81  & 0.60  & \nodata & North & \nodata \\
16 & 80.8801  & 0.3061  & -2.88  & 2.26  $\pm$ 0.38  & 16.65  $\pm$ 2.61  & 0.968  $\pm$ 0.097  & 0.51  & 3.35 & 0.88  & 1.28  & 1.80  & Y & North & \nodata \\
17 & 80.8612  & 0.4196  & -2.88  & 2.34  $\pm$ 0.44  & 18.41  $\pm$ 2.91  & 0.802  $\pm$ 0.080  & 0.50  & 2.96 & 0.85  & 1.35  & 1.92  & \nodata & North & \nodata \\
18 & 78.3636  & -1.0811  & -3.12  & 2.32  $\pm$ 0.51  & 24.14  $\pm$ 3.78  & 0.613  $\pm$ 0.061  & 0.48  & 2.71 & 0.89  & 1.11  & 1.25  & Y & South & \nodata \\
19 & 80.8990  & -0.2993  & -2.62  & 2.22  $\pm$ 0.39  & 24.39  $\pm$ 3.72  & 0.824  $\pm$ 0.082  & 0.49  & 3.68 & 1.10  & 1.04  & 1.14  & Y & North & \nodata \\
20 & 79.2722  & -0.1416  & -2.62  & 1.95  $\pm$ 0.36  & 7.72  $\pm$ 1.43  & 0.714  $\pm$ 0.071  & 0.56  & 1.69 & 0.34  & 0.67  & 0.56  & Y & South & \nodata \\
21 & 81.7692  & 0.6718  & -3.38  & 2.69  $\pm$ 0.39  & 25.39  $\pm$ 3.88  & 0.350  $\pm$ 0.035  & 0.33  & 1.61 & 1.66  & 0.42  & 0.14  & Y & North & \nodata \\
22 & 81.9079  & 0.8547  & -2.88  & 1.85  $\pm$ 0.34  & 20.88  $\pm$ 3.19  & 0.546  $\pm$ 0.055  & 0.57  & 2.19 & 0.42  & 0.58  & 0.43  & Y & North & \nodata \\
23 & 81.8260  & 0.8168  & -2.88  & 2.69  $\pm$ 0.39  & 21.51  $\pm$ 3.29  & 0.619  $\pm$ 0.062  & 0.42  & 2.53 & 1.24  & 0.96  & 0.82  & Y & North & \nodata \\
24 & 79.3100  & -0.3749  & -2.62  & 1.85  $\pm$ 0.42  & 21.94  $\pm$ 3.39  & 0.166  $\pm$ 0.017  & 0.31  & 0.69 & 0.79  & 0.44  & 0.14  & Y & South & \nodata \\
25 & 78.6605  & 0.1863  & -1.88  & 2.40  $\pm$ 0.39  & 14.20  $\pm$ 2.28  & 1.494  $\pm$ 0.149  & 0.64  & 4.71 & 0.64  & 1.18  & 1.88  & Y & South & \nodata \\
26 & 81.6746  & 0.5394  & -1.88  & 3.50  $\pm$ 0.43  & 36.48  $\pm$ 5.51  & 1.507  $\pm$ 0.151  & 0.44  & 9.11 & 3.72  & 1.11  & 1.16  & Y & North & \nodata \\
27 & 81.7188  & 0.5709  & -2.62  & 3.95  $\pm$ 0.49  & 35.61  $\pm$ 5.37  & 0.385  $\pm$ 0.039  & 0.25  & 2.28 & 5.46  & 0.62  & 0.21  & Y & North & \nodata \\
28 & 79.0892  & 0.4889  & -2.62  & 1.79  $\pm$ 0.36  & 17.50  $\pm$ 2.72  & 0.153  $\pm$ 0.015  & 0.29  & 0.55 & 0.81  & 0.46  & 0.14  & Y & South & \nodata \\
29 & 81.8386  & 0.8799  & -1.88  & 4.41  $\pm$ 0.52  & 33.21  $\pm$ 5.04  & 3.217  $\pm$ 0.322  & 0.72  & 18.10 & 1.71  & 1.50  & 3.46  & Y & North & \nodata \\
30 & 79.4358  & 1.0060  & -1.62  & 2.43  $\pm$ 0.45  & 9.13  $\pm$ 1.63  & 0.477  $\pm$ 0.048  & 0.48  & 1.21 & 0.40  & 0.75  & 0.58  & Y & South & \nodata \\
31 & 78.2435  & -1.3081  & -2.38  & 2.27  $\pm$ 0.63  & 19.73  $\pm$ 3.15  & 0.256  $\pm$ 0.026  & 0.33  & 0.99 & 1.02  & 0.68  & 0.33  & \nodata & DR13S & Y \\
32 & 81.4350  & 0.7096  & -1.62  & 2.43  $\pm$ 0.35  & 34.82  $\pm$ 5.28  & 1.399  $\pm$ 0.140  & 0.60  & 8.15 & 1.33  & 0.92  & 1.09  & Y & North & \nodata \\
33 & 77.8779  & -1.1253  & -2.12  & 2.02  $\pm$ 0.48  & 13.32  $\pm$ 2.23  & 0.173  $\pm$ 0.017  & 0.31  & 0.53 & 0.61  & 0.64  & 0.28  & Y & DR13S & \nodata \\
34 & 80.9116  & 0.3313  & -2.12  & 2.01  $\pm$ 0.36  & 17.04  $\pm$ 2.67  & 0.176  $\pm$ 0.018  & 0.29  & 0.62 & 0.92  & 0.57  & 0.21  & Y & DR13S & \nodata \\
35 & 79.7511  & 0.9934  & -0.88  & 2.20  $\pm$ 0.43  & 9.29  $\pm$ 1.59  & 0.857  $\pm$ 0.086  & 0.51  & 2.18 & 0.57  & 1.00  & 1.10  & Y & South & \nodata \\
36 & 78.2186  & -1.0496  & -1.62  & 1.77  $\pm$ 0.48  & 21.24  $\pm$ 3.35  & 0.291  $\pm$ 0.029  & 0.44  & 1.18 & 0.51  & 0.41  & 0.17  & \nodata & DR13S & \nodata \\
37 & 80.8612  & 0.3565  & -1.38  & 2.06  $\pm$ 0.41  & 21.54  $\pm$ 3.33  & 0.434  $\pm$ 0.043  & 0.38  & 1.78 & 1.16  & 1.25  & 1.26  & Y & North & \nodata \\
38 & 78.3004  & -1.1568  & -1.62  & 1.88  $\pm$ 0.56  & 17.33  $\pm$ 2.77  & 0.154  $\pm$ 0.015  & 0.34  & 0.55 & 0.51  & 0.40  & 0.13  & \nodata & DR13S & \nodata \\
39 & 77.9094  & -1.1631  & -0.62  & 2.28  $\pm$ 0.55  & 14.84  $\pm$ 2.42  & 1.610  $\pm$ 0.161  & 0.81  & 5.20 & 0.35  & 1.37  & 3.22  & Y & DR13S & \nodata \\
40 & 78.7735  & -1.0874  & -0.88  & 2.26  $\pm$ 0.44  & 24.28  $\pm$ 3.79  & 0.740  $\pm$ 0.074  & 0.60  & 3.29 & 0.54  & 0.63  & 0.54  & Y & South & \nodata \\
41 & 79.3478  & -0.2993  & -1.12  & 2.00  $\pm$ 0.41  & 9.99  $\pm$ 1.71  & 0.311  $\pm$ 0.031  & 0.30  & 0.82 & 1.06  & 0.88  & 0.51  & Y & South & \nodata \\
42 & 78.2498  & -1.3081  & -1.12  & 2.03  $\pm$ 0.63  & 16.96  $\pm$ 2.78  & 0.188  $\pm$ 0.019  & 0.35  & 0.66 & 0.56  & 0.42  & 0.14  & Y & DR13S & Y \\
43 & 77.8778  & -1.2009  & -1.12  & 2.26  $\pm$ 0.47  & 15.33  $\pm$ 2.49  & 0.165  $\pm$ 0.017  & 0.31  & 0.54 & 0.63  & 0.42  & 0.13  & \nodata & DR13S & \nodata \\
44 & 78.2752  & -1.1505  & -1.38  & 1.84  $\pm$ 0.54  & 17.00  $\pm$ 2.76  & 0.500  $\pm$ 0.050  & 0.56  & 1.75 & 0.36  & 0.88  & 0.93  & \nodata & DR13S & \nodata \\
45 & 78.0422  & -0.9046  & -1.12  & 1.92  $\pm$ 0.46  & 18.67  $\pm$ 2.93  & 0.373  $\pm$ 0.037  & 0.42  & 1.39 & 0.68  & 0.88  & 0.70  & \nodata & DR13S & \nodata \\
46 & 78.8497  & -0.3875  & -1.12  & 1.83  $\pm$ 0.35  & 19.79  $\pm$ 3.08  & 0.182  $\pm$ 0.018  & 0.29  & 0.70 & 1.04  & 0.49  & 0.16  & Y & South & \nodata \\
47 & 81.3530  & 0.5709  & -1.62  & 1.84  $\pm$ 0.35  & 29.87  $\pm$ 4.56  & 0.159  $\pm$ 0.016  & 0.33  & 0.83 & 0.85  & 0.44  & 0.15  & \nodata & North & \nodata \\
48 & 78.2246  & -1.3081  & -1.12  & 1.83  $\pm$ 0.53  & 13.65  $\pm$ 2.27  & 0.141  $\pm$ 0.014  & 0.35  & 0.43 & 0.37  & 0.30  & 0.08  & \nodata & DR13S & Y \\
49 & 78.2813  & -1.2892  & -1.12  & 1.91  $\pm$ 0.47  & 17.03  $\pm$ 2.75  & 0.176  $\pm$ 0.018  & 0.35  & 0.62 & 0.52  & 0.26  & 0.06  & Y & DR13S & Y \\
50 & 78.1053  & -0.7848  & -0.88  & 1.93  $\pm$ 0.51  & 26.34  $\pm$ 4.08  & 0.334  $\pm$ 0.033  & 0.51  & 1.58 & 0.43  & 0.37  & 0.17  & \nodata & DR13S & \nodata \\
51 & 78.0422  & -0.8857  & -0.62  & 1.92  $\pm$ 0.50  & 18.66  $\pm$ 2.94  & 0.364  $\pm$ 0.036  & 0.47  & 1.35 & 0.47  & 0.47  & 0.24  & \nodata & DR13S & \nodata \\
52 & 79.3226  & -0.1479  & -0.62  & 1.84  $\pm$ 0.37  & 7.89  $\pm$ 1.41  & 0.460  $\pm$ 0.046  & 0.54  & 1.09 & 0.24  & 0.50  & 0.31  & Y & South & \nodata \\
53 & 80.9116  & 0.3187  & -0.88  & 1.80  $\pm$ 0.38  & 18.83  $\pm$ 2.93  & 0.130  $\pm$ 0.013  & 0.31  & 0.49 & 0.56  & 0.28  & 0.06  & Y & North & \nodata \\
54 & 80.6972  & 0.6970  & -1.12  & 2.00  $\pm$ 0.42  & 19.81  $\pm$ 3.10  & 0.590  $\pm$ 0.059  & 0.52  & 2.28 & 0.57  & 0.70  & 0.57  & Y & North & \nodata \\
55 & 81.8575  & 0.9619  & -0.88  & 2.01  $\pm$ 0.35  & 19.93  $\pm$ 3.10  & 0.194  $\pm$ 0.019  & 0.36  & 0.75 & 0.58  & 0.30  & 0.08  & Y & North & \nodata \\
56 & 77.9033  & -1.0433  & -0.62  & 1.79  $\pm$ 0.53  & 9.31  $\pm$ 1.72  & 0.136  $\pm$ 0.014  & 0.33  & 0.35 & 0.36  & 0.29  & 0.07  & \nodata & DR13S & \nodata \\
57 & 78.0737  & -0.8668  & -0.38  & 2.68  $\pm$ 0.53  & 20.73  $\pm$ 3.25  & 0.809  $\pm$ 0.081  & 0.52  & 3.23 & 0.81  & 1.28  & 1.81  & Y & DR13S & \nodata \\
58 & 78.0928  & -0.6713  & -0.12  & 3.23  $\pm$ 0.60  & 21.00  $\pm$ 3.31  & 0.581  $\pm$ 0.058  & 0.37  & 2.34 & 1.66  & 0.69  & 0.39  & Y & DR13S & \nodata \\
59 & 78.1118  & -0.6334  & -0.38  & 3.25  $\pm$ 0.52  & 23.88  $\pm$ 3.72  & 0.607  $\pm$ 0.061  & 0.37  & 2.67 & 1.89  & 0.70  & 0.39  & Y & DR13S & \nodata \\
60 & 79.2784  & 0.3817  & 1.38  & 2.28  $\pm$ 0.39  & 13.29  $\pm$ 2.12  & 0.893  $\pm$ 0.089  & 0.50  & 2.71 & 0.78  & 2.16  & 4.92  & Y & South & \nodata \\
61 & 79.2406  & 0.5268  & 0.38  & 3.76  $\pm$ 0.49  & 9.69  $\pm$ 1.62  & 5.040  $\pm$ 0.504  & 0.97  & 13.10 & 0.51  & 1.05  & 2.30  & Y & South & \nodata \\
62 & 77.8715  & -1.2073  & 0.12  & 1.99  $\pm$ 0.52  & 15.89  $\pm$ 2.56  & 0.293  $\pm$ 0.029  & 0.39  & 0.99 & 0.60  & 0.54  & 0.26  & \nodata & DR13S & \nodata \\
63 & 79.5181  & 0.0034  & 0.12  & 1.91  $\pm$ 0.39  & 8.86  $\pm$ 1.58  & 0.798  $\pm$ 0.080  & 0.68  & 1.99 & 0.23  & 1.07  & 1.66  & Y & South & \nodata \\
64 & 79.4613  & 0.1989  & 0.12  & 2.57  $\pm$ 0.40  & 9.45  $\pm$ 1.61  & 2.137  $\pm$ 0.214  & 0.97  & 5.48 & 0.22  & 0.73  & 1.11  & Y & South & \nodata \\
65 & 79.4929  & 0.2619  & -0.12  & 1.83  $\pm$ 0.33  & 7.19  $\pm$ 1.28  & 0.641  $\pm$ 0.064  & 0.54  & 1.48 & 0.33  & 0.99  & 1.14  & \nodata & South & \nodata \\
66 & 77.8273  & -1.2640  & -0.12  & 2.13  $\pm$ 0.50  & 16.44  $\pm$ 2.63  & 0.505  $\pm$ 0.051  & 0.54  & 1.73 & 0.39  & 1.36  & 2.14  & \nodata & DR13S & \nodata \\
67 & 78.1367  & -0.9046  & 0.62  & 3.48  $\pm$ 0.51  & 23.77  $\pm$ 3.70  & 4.214  $\pm$ 0.421  & 0.93  & 18.50 & 0.81  & 1.34  & 3.54  & Y & DR13S & \nodata \\
68 & 78.9506  & -0.1858  & 1.62  & 2.41  $\pm$ 0.41  & 13.25  $\pm$ 2.14  & 0.717  $\pm$ 0.072  & 0.50  & 2.17 & 0.62  & 1.38  & 2.01  & Y & South & \nodata \\
69 & 79.3226  & 0.3313  & 1.38  & 3.85  $\pm$ 0.51  & 15.23  $\pm$ 2.37  & 2.975  $\pm$ 0.298  & 0.60  & 9.76 & 1.64  & 1.28  & 2.08  & Y & South & \nodata \\
70 & 79.1712  & 0.5898  & 0.12  & 2.26  $\pm$ 0.41  & 8.55  $\pm$ 1.50  & 1.092  $\pm$ 0.109  & 0.76  & 2.68 & 0.22  & 0.53  & 0.49  & Y & South & \nodata \\
71 & 81.8701  & 1.0375  & 0.12  & 2.40  $\pm$ 0.38  & 24.16  $\pm$ 3.75  & 1.019  $\pm$ 0.102  & 0.63  & 4.52 & 0.65  & 0.70  & 0.67  & Y & North & \nodata \\
72 & 78.0549  & -0.7217  & 0.38  & 1.94  $\pm$ 0.46  & 18.57  $\pm$ 3.03  & 0.193  $\pm$ 0.019  & 0.37  & 0.72 & 0.51  & 0.44  & 0.16  & \nodata & DR13S & \nodata \\
73 & 78.1622  & -0.5578  & 0.12  & 1.83  $\pm$ 0.45  & 24.34  $\pm$ 3.78  & 0.190  $\pm$ 0.019  & 0.42  & 0.84 & 0.41  & 0.07  & 0.02  & Y & DR13S & \nodata \\
74 & 79.3100  & -0.1668  & 0.12  & 2.03  $\pm$ 0.36  & 8.87  $\pm$ 1.53  & 0.456  $\pm$ 0.046  & 0.48  & 1.14 & 0.38  & 0.80  & 0.66  & Y & South & \nodata \\
75 & 79.7829  & 0.0728  & 0.12  & 2.09  $\pm$ 0.44  & 7.66  $\pm$ 1.44  & 0.415  $\pm$ 0.042  & 0.49  & 0.98 & 0.29  & 0.49  & 0.27  & \nodata & South & \nodata \\
76 & 79.1144  & 0.6466  & -0.12  & 2.10  $\pm$ 0.39  & 8.83  $\pm$ 1.54  & 0.149  $\pm$ 0.015  & 0.36  & 0.37 & 0.29  & 0.10  & 0.02  & Y & South & \nodata \\
77 & 78.9630  & 0.7412  & 0.12  & 1.84  $\pm$ 0.48  & 8.27  $\pm$ 1.63  & 0.128  $\pm$ 0.013  & 0.34  & 0.31 & 0.29  & 0.18  & 0.04  & \nodata & South & \nodata \\
78 & 79.4046  & 0.1421  & 0.38  & 1.87  $\pm$ 0.38  & 6.88  $\pm$ 1.30  & 0.275  $\pm$ 0.027  & 0.40  & 0.62 & 0.35  & 0.40  & 0.15  & \nodata & South & \nodata \\
79 & 79.4109  & 0.2619  & 0.62  & 1.83  $\pm$ 0.35  & 5.70  $\pm$ 1.17  & 0.354  $\pm$ 0.035  & 0.44  & 0.76 & 0.31  & 0.51  & 0.26  & Y & South & \nodata \\
80 & 79.2532  & 0.4007  & 0.38  & 2.19  $\pm$ 0.44  & 8.91  $\pm$ 1.52  & 0.215  $\pm$ 0.022  & 0.30  & 0.54 & 0.70  & 0.54  & 0.20  & Y & South & \nodata \\
81 & 78.9251  & 0.7853  & 0.62  & 2.58  $\pm$ 0.44  & 10.49  $\pm$ 1.83  & 1.218  $\pm$ 0.122  & 0.73  & 3.27 & 0.30  & 0.52  & 0.45  & Y & South & \nodata \\
82 & 81.7124  & 1.6807  & 0.38  & 1.83  $\pm$ 0.45  & 20.75  $\pm$ 3.26  & 0.168  $\pm$ 0.017  & 0.30  & 0.67 & 0.87  & 0.78  & 0.40  & \nodata & North & Y \\
83 & 78.1620  & -0.8415  & 0.62  & 2.12  $\pm$ 0.49  & 18.50  $\pm$ 2.89  & 0.275  $\pm$ 0.028  & 0.36  & 1.02 & 0.79  & 0.51  & 0.21  & \nodata & DR13S & \nodata \\
84 & 79.7073  & 0.1295  & 0.88  & 2.32  $\pm$ 0.40  & 8.85  $\pm$ 1.47  & 0.968  $\pm$ 0.097  & 0.74  & 2.41 & 0.21  & 0.46  & 0.37  & Y & South & \nodata \\
85 & 77.8968  & -1.1946  & 1.38  & 2.00  $\pm$ 0.51  & 15.22  $\pm$ 2.50  & 0.626  $\pm$ 0.063  & 0.59  & 2.05 & 0.36  & 0.72  & 0.66  & Y & DR13S & \nodata \\
86 & 79.3667  & 0.3817  & 0.88  & 3.14  $\pm$ 0.43  & 11.75  $\pm$ 1.88  & 0.255  $\pm$ 0.026  & 0.30  & 0.73 & 0.94  & 0.28  & 0.06  & \nodata & South & \nodata \\
87 & 79.2532  & 0.4448  & 1.12  & 2.82  $\pm$ 0.42  & 9.13  $\pm$ 1.55  & 0.633  $\pm$ 0.063  & 0.48  & 1.60 & 0.53  & 0.65  & 0.44  & Y & South & \nodata \\
88 & 80.0225  & -0.4254  & 1.38  & 2.30  $\pm$ 0.40  & 9.24  $\pm$ 1.64  & 0.540  $\pm$ 0.054  & 0.46  & 1.37 & 0.50  & 0.62  & 0.40  & Y & South & \nodata \\
89 & 81.8827  & 1.1321  & 1.88  & 2.08  $\pm$ 0.44  & 11.50  $\pm$ 1.95  & 0.599  $\pm$ 0.060  & 0.56  & 1.69 & 0.34  & 0.76  & 0.71  & \nodata & North & \nodata \\
90 & 79.3415  & -0.4695  & 1.62  & 1.94  $\pm$ 0.44  & 8.59  $\pm$ 1.56  & 0.517  $\pm$ 0.052  & 0.52  & 1.27 & 0.32  & 0.80  & 0.72  & Y & South & \nodata \\
91 & 79.8901  & -0.2236  & 1.38  & 2.06  $\pm$ 0.41  & 9.89  $\pm$ 1.74  & 0.173  $\pm$ 0.017  & 0.34  & 0.45 & 0.42  & 0.30  & 0.08  & \nodata & South & \nodata \\
92 & 79.2406  & 0.4196  & 1.88  & 3.09  $\pm$ 0.45  & 11.56  $\pm$ 1.91  & 1.216  $\pm$ 0.122  & 0.56  & 3.43 & 0.68  & 0.74  & 0.67  & Y & South & \nodata \\
93 & 78.0737  & -0.8920  & 1.38  & 2.36  $\pm$ 0.53  & 16.34  $\pm$ 2.61  & 0.175  $\pm$ 0.018  & 0.34  & 0.60 & 0.56  & 0.25  & 0.06  & \nodata & DR13S & \nodata \\
94 & 77.8208  & -1.3586  & 3.38  & 2.06  $\pm$ 0.54  & 17.29  $\pm$ 2.78  & 1.513  $\pm$ 0.151  & 0.84  & 5.36 & 0.32  & 1.20  & 2.57  & Y & DR13S & \nodata \\
95 & 78.5531  & 0.6907  & 2.12  & 1.83  $\pm$ 0.44  & 10.68  $\pm$ 1.88  & 0.166  $\pm$ 0.017  & 0.30  & 0.45 & 0.59  & 0.25  & 0.05  & \nodata & South & \nodata \\
96 & 81.3845  & 0.8862  & 2.88  & 1.78  $\pm$ 0.34  & 17.44  $\pm$ 2.71  & 0.237  $\pm$ 0.024  & 0.35  & 0.85 & 0.72  & 0.51  & 0.21  & Y & North & \nodata \\
97 & 77.9850  & -1.2577  & 3.12  & 1.78  $\pm$ 0.50  & 10.34  $\pm$ 1.81  & 0.391  $\pm$ 0.039  & 0.47  & 1.04 & 0.36  & 0.86  & 0.75  & Y & DR13S & \nodata \\
98 & 79.2406  & 0.2367  & 2.88  & 1.84  $\pm$ 0.40  & 27.47  $\pm$ 4.19  & 0.308  $\pm$ 0.031  & 0.38  & 1.50 & 0.98  & 1.06  & 0.91  & \nodata & South & \nodata \\
99 & 81.7061  & 1.2141  & 2.88  & 2.22  $\pm$ 0.39  & 11.50  $\pm$ 1.98  & 0.166  $\pm$ 0.017  & 0.31  & 0.47 & 0.54  & 0.34  & 0.09  & \nodata & North & \nodata \\
100 & 81.7818  & 1.2519  & 2.88  & 2.05  $\pm$ 0.39  & 16.18  $\pm$ 2.62  & 0.190  $\pm$ 0.019  & 0.33  & 0.65 & 0.67  & 0.55  & 0.23  & \nodata & North & \nodata \\
101 & 77.9470  & -1.3460  & 3.12  & 1.87  $\pm$ 0.52  & 15.73  $\pm$ 2.61  & 0.327  $\pm$ 0.033  & 0.44  & 1.09 & 0.47  & 0.64  & 0.40  & \nodata & DR13S & \nodata \\
102 & 79.4864  & -0.6587  & 3.38  & 1.90  $\pm$ 0.33  & 12.69  $\pm$ 2.11  & 0.354  $\pm$ 0.035  & 0.45  & 1.05 & 0.41  & 0.55  & 0.31  & Y & South & \nodata \\
103 & 81.3530  & 0.3817  & 3.12  & 1.84  $\pm$ 0.33  & 10.16  $\pm$ 1.67  & 0.215  $\pm$ 0.022  & 0.45  & 0.57 & 0.22  & 0.01  & 0.01  & \nodata & North & \nodata \\
104 & 81.2773  & 0.5394  & 3.12  & 2.13  $\pm$ 0.32  & 10.41  $\pm$ 1.67  & 1.354  $\pm$ 0.135  & 0.86  & 3.63 & 0.20  & 1.01  & 1.88  & \nodata & North & \nodata \\
105 & 77.9660  & -1.3144  & 3.88  & 2.52  $\pm$ 0.51  & 12.73  $\pm$ 2.21  & 0.325  $\pm$ 0.033  & 0.39  & 0.96 & 0.58  & 0.67  & 0.38  & Y & DR13S & \nodata \\
106 & 81.3845  & 0.4133  & 3.88  & 1.91  $\pm$ 0.34  & 11.42  $\pm$ 1.83  & 0.648  $\pm$ 0.065  & 0.52  & 1.82 & 0.46  & 0.57  & 0.38  & Y & North & \nodata \\
107 & 81.7629  & 1.3528  & 3.88  & 3.65  $\pm$ 0.50  & 11.21  $\pm$ 1.90  & 2.011  $\pm$ 0.201  & 0.76  & 5.59 & 0.45  & 1.35  & 2.96  & Y & North & \nodata \\
108 & 81.7314  & 1.2961  & 3.62  & 2.22  $\pm$ 0.39  & 13.84  $\pm$ 2.26  & 0.881  $\pm$ 0.088  & 0.57  & 2.74 & 0.53  & 0.69  & 0.59  & \nodata & North & \nodata \\
109 & 81.6998  & 1.3528  & 3.62  & 2.06  $\pm$ 0.40  & 10.02  $\pm$ 1.73  & 0.232  $\pm$ 0.023  & 0.29  & 0.61 & 0.91  & 0.47  & 0.14  & \nodata & North & \nodata \\
110 & 79.3036  & -0.6713  & 4.12  & 3.39  $\pm$ 0.49  & 15.93  $\pm$ 2.63  & 1.521  $\pm$ 0.152  & 0.76  & 5.13 & 0.41  & 0.79  & 1.04  & Y & South & \nodata \\
111 & 77.9849  & -1.2829  & 3.88  & 2.12  $\pm$ 0.55  & 13.90  $\pm$ 2.30  & 0.175  $\pm$ 0.018  & 0.35  & 0.54 & 0.46  & 0.20  & 0.04  & \nodata & DR13S & \nodata \\
112 & 78.9442  & -0.5515  & 4.38  & 1.91  $\pm$ 0.33  & 8.63  $\pm$ 1.60  & 0.149  $\pm$ 0.015  & 0.34  & 0.37 & 0.34  & 0.31  & 0.08  & Y & South & \nodata \\
113 & 81.9521  & -0.3308  & 4.62  & 2.04  $\pm$ 0.63  & 8.65  $\pm$ 2.45  & 0.217  $\pm$ 0.022  & 0.37  & 0.54 & 0.38  & 0.42  & 0.15  & \nodata & North & Y \\
114 & 78.9695  & 0.3565  & 5.38  & 2.98  $\pm$ 0.43  & 27.45  $\pm$ 4.20  & 2.278  $\pm$ 0.228  & 0.70  & 11.10 & 1.14  & 1.59  & 3.76  & Y & South & \nodata \\
115 & 79.5557  & -0.8037  & 5.12  & 3.23  $\pm$ 0.47  & 13.98  $\pm$ 2.27  & 1.617  $\pm$ 0.162  & 0.60  & 5.05 & 0.85  & 1.20  & 1.83  & Y & South & \nodata \\
116 & 80.8927  & -0.3308  & 5.12  & 2.06  $\pm$ 0.39  & 8.95  $\pm$ 1.59  & 1.163  $\pm$ 0.116  & 0.80  & 2.91 & 0.20  & 0.57  & 0.57  & \nodata & North & \nodata \\
117 & 80.8423  & -0.1921  & 5.88  & 1.81  $\pm$ 0.35  & 8.71  $\pm$ 1.53  & 0.944  $\pm$ 0.094  & 0.55  & 2.34 & 0.50  & 1.50  & 2.60  & Y & North & \nodata \\
118 & 81.0756  & -0.4758  & 5.62  & 2.49  $\pm$ 0.39  & 11.47  $\pm$ 1.89  & 0.514  $\pm$ 0.051  & 0.50  & 1.44 & 0.41  & 0.52  & 0.31  & Y & North & \nodata \\
119 & 81.0062  & -0.3749  & 5.12  & 1.89  $\pm$ 0.40  & 8.18  $\pm$ 1.45  & 0.195  $\pm$ 0.020  & 0.35  & 0.47 & 0.40  & 0.25  & 0.06  & \nodata & North & \nodata \\
120 & 81.2837  & -0.2740  & 5.38  & 2.01  $\pm$ 0.40  & 6.81  $\pm$ 1.33  & 0.491  $\pm$ 0.049  & 0.52  & 1.11 & 0.28  & 0.55  & 0.35  & \nodata & North & \nodata \\
121 & 79.0324  & 0.6592  & 6.88  & 3.14  $\pm$ 0.46  & 31.35  $\pm$ 4.77  & 2.735  $\pm$ 0.274  & 0.73  & 14.70 & 1.33  & 1.44  & 3.20  & Y & South & \nodata \\
122 & 81.0503  & -0.5199  & 5.88  & 2.10  $\pm$ 0.39  & 10.59  $\pm$ 1.72  & 0.261  $\pm$ 0.026  & 0.34  & 0.71 & 0.66  & 0.76  & 0.42  & \nodata & North & \nodata \\
123 & 79.1334  & -0.3749  & 6.62  & 1.91  $\pm$ 0.36  & 24.84  $\pm$ 3.80  & 0.338  $\pm$ 0.034  & 0.35  & 1.53 & 1.29  & 1.17  & 1.01  & Y & South & \nodata \\
124 & 81.1954  & -0.2866  & 5.38  & 2.58  $\pm$ 0.44  & 7.88  $\pm$ 1.45  & 0.694  $\pm$ 0.069  & 0.58  & 1.65 & 0.30  & 0.49  & 0.32  & Y & North & \nodata \\
125 & 81.0251  & -0.4506  & 5.88  & 2.60  $\pm$ 0.41  & 11.05  $\pm$ 1.84  & 1.832  $\pm$ 0.183  & 0.87  & 5.05 & 0.27  & 0.51  & 0.52  & Y & North & \nodata \\
126 & 80.9873  & -0.4064  & 5.88  & 2.27  $\pm$ 0.38  & 8.67  $\pm$ 1.54  & 0.195  $\pm$ 0.020  & 0.39  & 0.48 & 0.29  & 0.23  & 0.06  & \nodata & North & \nodata \\
127 & 81.0504  & 0.2998  & 5.62  & 1.93  $\pm$ 0.39  & 10.44  $\pm$ 1.79  & 0.147  $\pm$ 0.015  & 0.36  & 0.39 & 0.30  & 0.18  & 0.04  & \nodata & North & \nodata \\
128 & 79.7325  & -0.2614  & 6.38  & 2.09  $\pm$ 0.40  & 11.15  $\pm$ 1.89  & 0.386  $\pm$ 0.039  & 0.41  & 1.07 & 0.56  & 0.82  & 0.59  & Y & South & \nodata \\
129 & 78.6164  & -0.1038  & 6.38  & 2.00  $\pm$ 0.45  & 8.44  $\pm$ 1.63  & 0.215  $\pm$ 0.022  & 0.38  & 0.53 & 0.34  & 0.35  & 0.11  & \nodata & South & \nodata \\
130 & 78.9631  & 0.5331  & 6.38  & 2.41  $\pm$ 0.44  & 21.47  $\pm$ 3.30  & 0.395  $\pm$ 0.040  & 0.36  & 1.61 & 1.25  & 0.84  & 0.55  & Y & South & \nodata \\
131 & 78.9757  & 0.6087  & 6.38  & 1.93  $\pm$ 0.41  & 19.34  $\pm$ 3.04  & 0.183  $\pm$ 0.018  & 0.30  & 0.70 & 0.91  & 0.43  & 0.13  & Y & South & \nodata \\
132 & 78.5786  & -0.0281  & 6.88  & 2.22  $\pm$ 0.40  & 8.69  $\pm$ 1.53  & 0.927  $\pm$ 0.093  & 0.72  & 2.29 & 0.22  & 0.51  & 0.43  & Y & South & \nodata \\
133 & 79.7577  & -0.1542  & 6.88  & 1.79  $\pm$ 0.36  & 11.29  $\pm$ 1.89  & 0.256  $\pm$ 0.026  & 0.38  & 0.72 & 0.47  & 0.76  & 0.48  & \nodata & South & \nodata \\
134 & 81.4539  & 0.4637  & 7.88  & 2.18  $\pm$ 0.36  & 40.61  $\pm$ 6.13  & 0.502  $\pm$ 0.050  & 0.47  & 3.31 & 1.15  & 0.93  & 0.87  & Y & North & \nodata \\
135 & 79.2214  & 0.9556  & 7.38  & 3.00  $\pm$ 0.49  & 28.31  $\pm$ 4.33  & 1.077  $\pm$ 0.108  & 0.58  & 5.36 & 0.99  & 0.79  & 0.78  & Y & South & \nodata \\
136 & 79.1773  & 0.9556  & 7.12  & 2.85  $\pm$ 0.47  & 26.65  $\pm$ 4.11  & 0.499  $\pm$ 0.050  & 0.39  & 2.38 & 1.44  & 0.70  & 0.42  & Y & South & \nodata \\
137 & 82.0277  & -0.2614  & 7.62  & 2.30  $\pm$ 0.40  & 11.29  $\pm$ 1.91  & 0.571  $\pm$ 0.057  & 0.47  & 1.59 & 0.55  & 0.57  & 0.34  & Y & North & Y \\
138 & 81.5296  & 0.2241  & 8.88  & 2.60  $\pm$ 0.36  & 24.69  $\pm$ 3.75  & 3.347  $\pm$ 0.335  & 0.78  & 15.10 & 1.14  & 1.68  & 4.65  & Y & North & \nodata \\
139 & 79.0764  & 0.9303  & 7.88  & 2.11  $\pm$ 0.43  & 30.59  $\pm$ 4.65  & 0.245  $\pm$ 0.025  & 0.31  & 1.29 & 1.49  & 0.73  & 0.37  & Y & South & \nodata \\
140 & 79.0261  & 0.6214  & 8.12  & 1.92  $\pm$ 0.37  & 19.46  $\pm$ 3.01  & 0.150  $\pm$ 0.015  & 0.29  & 0.57 & 0.85  & 0.65  & 0.27  & \nodata & South & \nodata \\
141 & 79.1205  & 0.9556  & 7.62  & 2.16  $\pm$ 0.42  & 30.58  $\pm$ 4.67  & 0.661  $\pm$ 0.066  & 0.47  & 3.48 & 1.21  & 0.73  & 0.55  & Y & South & \nodata \\
142 & 81.4161  & 0.4196  & 8.12  & 2.38  $\pm$ 0.38  & 27.91  $\pm$ 4.23  & 0.594  $\pm$ 0.059  & 0.44  & 2.92 & 1.27  & 0.61  & 0.36  & \nodata & North & \nodata \\
143 & 80.3693  & 0.4448  & 8.62  & 2.45  $\pm$ 0.41  & 35.86  $\pm$ 5.46  & 0.810  $\pm$ 0.081  & 0.44  & 4.83 & 1.97  & 0.94  & 0.85  & Y & North & \nodata \\
144 & 81.6494  & 0.7222  & 8.38  & 2.06  $\pm$ 0.35  & 24.10  $\pm$ 3.68  & 0.367  $\pm$ 0.037  & 0.34  & 1.63 & 1.52  & 0.56  & 0.23  & Y & North & \nodata \\
145 & 78.3768  & 0.3187  & 8.38  & 2.38  $\pm$ 0.47  & 12.54  $\pm$ 2.18  & 0.533  $\pm$ 0.053  & 0.54  & 1.57 & 0.36  & 0.50  & 0.31  & Y & South & \nodata \\
146 & 81.8701  & 0.7790  & 9.62  & 2.76  $\pm$ 0.37  & 24.96  $\pm$ 3.81  & 3.070  $\pm$ 0.307  & 0.72  & 13.90 & 1.35  & 1.48  & 3.32  & Y & North & \nodata \\
147 & 81.7061  & -0.0344  & 8.88  & 2.82  $\pm$ 0.38  & 21.94  $\pm$ 3.38  & 0.475  $\pm$ 0.048  & 0.42  & 1.97 & 0.97  & 0.43  & 0.18  & \nodata & North & \nodata \\
148 & 81.6872  & -0.0218  & 9.38  & 2.42  $\pm$ 0.36  & 30.22  $\pm$ 4.58  & 0.369  $\pm$ 0.037  & 0.45  & 1.93 & 0.75  & 0.63  & 0.40  & Y & North & \nodata \\
149 & 78.3578  & 0.3187  & 8.88  & 1.94  $\pm$ 0.40  & 12.44  $\pm$ 2.18  & 0.828  $\pm$ 0.083  & 0.69  & 2.43 & 0.26  & 0.56  & 0.48  & Y & South & \nodata \\
150 & 81.4602  & 0.4637  & 9.38  & 2.29  $\pm$ 0.35  & 36.78  $\pm$ 5.55  & 0.324  $\pm$ 0.032  & 0.39  & 1.97 & 1.19  & 0.49  & 0.21  & Y & North & \nodata \\
151 & 81.5359  & 0.1043  & 9.38  & 2.22  $\pm$ 0.37  & 22.38  $\pm$ 3.43  & 0.454  $\pm$ 0.045  & 0.44  & 1.91 & 0.83  & 0.46  & 0.22  & Y & North & \nodata \\
152 & 78.3074  & 0.1674  & 9.38  & 2.34  $\pm$ 0.50  & 10.59  $\pm$ 1.87  & 0.210  $\pm$ 0.021  & 0.34  & 0.57 & 0.53  & 0.34  & 0.10  & \nodata & South & \nodata \\
153 & 78.3579  & 0.1989  & 9.12  & 2.15  $\pm$ 0.44  & 13.31  $\pm$ 2.25  & 0.570  $\pm$ 0.057  & 0.48  & 1.73 & 0.54  & 0.51  & 0.29  & \nodata & South & \nodata \\
154 & 78.4020  & 0.2556  & 9.12  & 2.24  $\pm$ 0.42  & 13.13  $\pm$ 2.13  & 1.601  $\pm$ 0.160  & 0.85  & 4.83 & 0.28  & 0.67  & 0.85  & Y & South & \nodata \\
155 & 82.2484  & 0.1674  & 9.38  & 2.34  $\pm$ 0.41  & 13.54  $\pm$ 2.20  & 0.254  $\pm$ 0.025  & 0.38  & 0.78 & 0.51  & 0.26  & 0.07  & Y & North & Y \\
156 & 82.2547  & 0.2052  & 9.38  & 2.99  $\pm$ 0.54  & 14.26  $\pm$ 2.51  & 0.306  $\pm$ 0.031  & 0.31  & 0.97 & 1.12  & 0.41  & 0.12  & \nodata & North & Y \\
157 & 82.1980  & 0.1232  & 10.12  & 2.04  $\pm$ 0.35  & 14.02  $\pm$ 2.31  & 0.801  $\pm$ 0.080  & 0.65  & 2.51 & 0.33  & 0.81  & 0.91  & Y & North & \nodata \\
158 & 78.2885  & 0.1737  & 10.12  & 2.57  $\pm$ 0.43  & 12.72  $\pm$ 2.10  & 0.510  $\pm$ 0.051  & 0.44  & 1.51 & 0.66  & 0.53  & 0.27  & \nodata & South & \nodata \\
159 & 81.8260  & 1.2645  & 10.62  & 2.02  $\pm$ 0.37  & 28.87  $\pm$ 4.44  & 0.792  $\pm$ 0.079  & 0.54  & 4.00 & 0.89  & 1.19  & 1.63  & Y & North & \nodata \\
160 & 82.2673  & 0.1547  & 9.88  & 1.93  $\pm$ 0.51  & 11.13  $\pm$ 2.18  & 0.148  $\pm$ 0.015  & 0.31  & 0.41 & 0.47  & 0.27  & 0.06  & \nodata & North & Y \\
161 & 78.1876  & 0.0980  & 10.62  & 2.63  $\pm$ 0.44  & 14.45  $\pm$ 2.38  & 0.390  $\pm$ 0.039  & 0.47  & 1.24 & 0.43  & 0.42  & 0.20  & Y & South & \nodata \\
162 & 78.2255  & 0.1737  & 10.38  & 2.21  $\pm$ 0.50  & 14.73  $\pm$ 2.41  & 0.211  $\pm$ 0.021  & 0.34  & 0.68 & 0.63  & 0.23  & 0.05  & Y & South & \nodata \\
163 & 79.3223  & -0.9866  & 11.38  & 1.89  $\pm$ 0.38  & 18.30  $\pm$ 2.93  & 0.124  $\pm$ 0.013  & 0.30  & 0.46 & 0.60  & 0.36  & 0.10  & Y & South & \nodata \\
164 & 81.9016  & 0.7916  & 11.62  & 2.17  $\pm$ 0.35  & 26.91  $\pm$ 4.10  & 0.637  $\pm$ 0.064  & 0.45  & 3.05 & 1.18  & 0.99  & 0.95  & Y & North & \nodata \\
165 & 81.8890  & 1.2393  & 11.88  & 1.89  $\pm$ 0.36  & 24.54  $\pm$ 3.80  & 0.141  $\pm$ 0.014  & 0.29  & 0.63 & 0.94  & 0.54  & 0.19  & Y & North & \nodata \\
166 & 79.2970  & -1.0244  & 12.12  & 2.33  $\pm$ 0.50  & 16.03  $\pm$ 2.61  & 0.533  $\pm$ 0.053  & 0.48  & 1.80 & 0.59  & 0.43  & 0.21  & Y & South & Y \\
167 & 81.1386  & 0.6970  & 12.12  & 1.90  $\pm$ 0.35  & 14.74  $\pm$ 2.34  & 0.521  $\pm$ 0.052  & 0.44  & 1.68 & 0.69  & 0.68  & 0.44  & Y & North & \nodata \\
168 & 81.8323  & 1.2015  & 13.12  & 3.74  $\pm$ 0.48  & 29.73  $\pm$ 4.54  & 1.851  $\pm$ 0.185  & 0.56  & 9.55 & 1.97  & 1.32  & 2.05  & Y & North & \nodata \\
169 & 81.2205  & 0.8988  & 13.62  & 2.73  $\pm$ 0.45  & 28.82  $\pm$ 4.43  & 2.346  $\pm$ 0.235  & 0.86  & 11.80 & 0.66  & 2.02  & 7.39  & Y & North & \nodata \\
170 & 81.2584  & 0.9871  & 13.62  & 2.01  $\pm$ 0.40  & 33.48  $\pm$ 5.12  & 0.706  $\pm$ 0.071  & 0.53  & 3.99 & 0.96  & 0.76  & 0.67  & Y & North & \nodata \\
171 & 81.2647  & 0.9114  & 13.88  & 2.27  $\pm$ 0.46  & 31.95  $\pm$ 4.90  & 0.512  $\pm$ 0.051  & 0.47  & 2.79 & 0.97  & 0.79  & 0.64  & Y & North & \nodata \\
172 & 81.3467  & 0.7412  & 15.62  & 2.24  $\pm$ 0.36  & 27.66  $\pm$ 4.21  & 0.603  $\pm$ 0.060  & 0.46  & 2.95 & 1.08  & 0.72  & 0.52  & Y & North & \nodata \\
173 & 81.1638  & 0.8042  & 15.88  & 1.98  $\pm$ 0.37  & 19.17  $\pm$ 3.05  & 0.223  $\pm$ 0.022  & 0.36  & 0.85 & 0.65  & 0.57  & 0.26  & \nodata & North & \nodata \\
174 & 81.2962  & 1.0564  & 16.12  & 1.82  $\pm$ 0.42  & 32.73  $\pm$ 5.01  & 0.166  $\pm$ 0.017  & 0.23  & 0.92 & 2.69  & 0.67  & 0.23  & Y & North & \nodata 
\enddata
\tablecomments{The columns give (1) object ID, (2) peak galactic longitude, (3) peak galactic latitude, (4) peak velocity, (5) peak C$^{18}$O main beam temperature, (6) peak $^{12}$CO main beam temperature, (7) C$^{18}$O integrated intensity, (8) core radius, (9) LTE mass, (10) mean density of hydrogen molecule, (11) intensity-weighted velocity FWHM, (12) virial mass, (13) existence of protostar, (14) region, and (15) map edge object or not.}
\end{deluxetable*}
\end{longrotatetable}

\begin{figure}
\epsscale{1}
\plotone{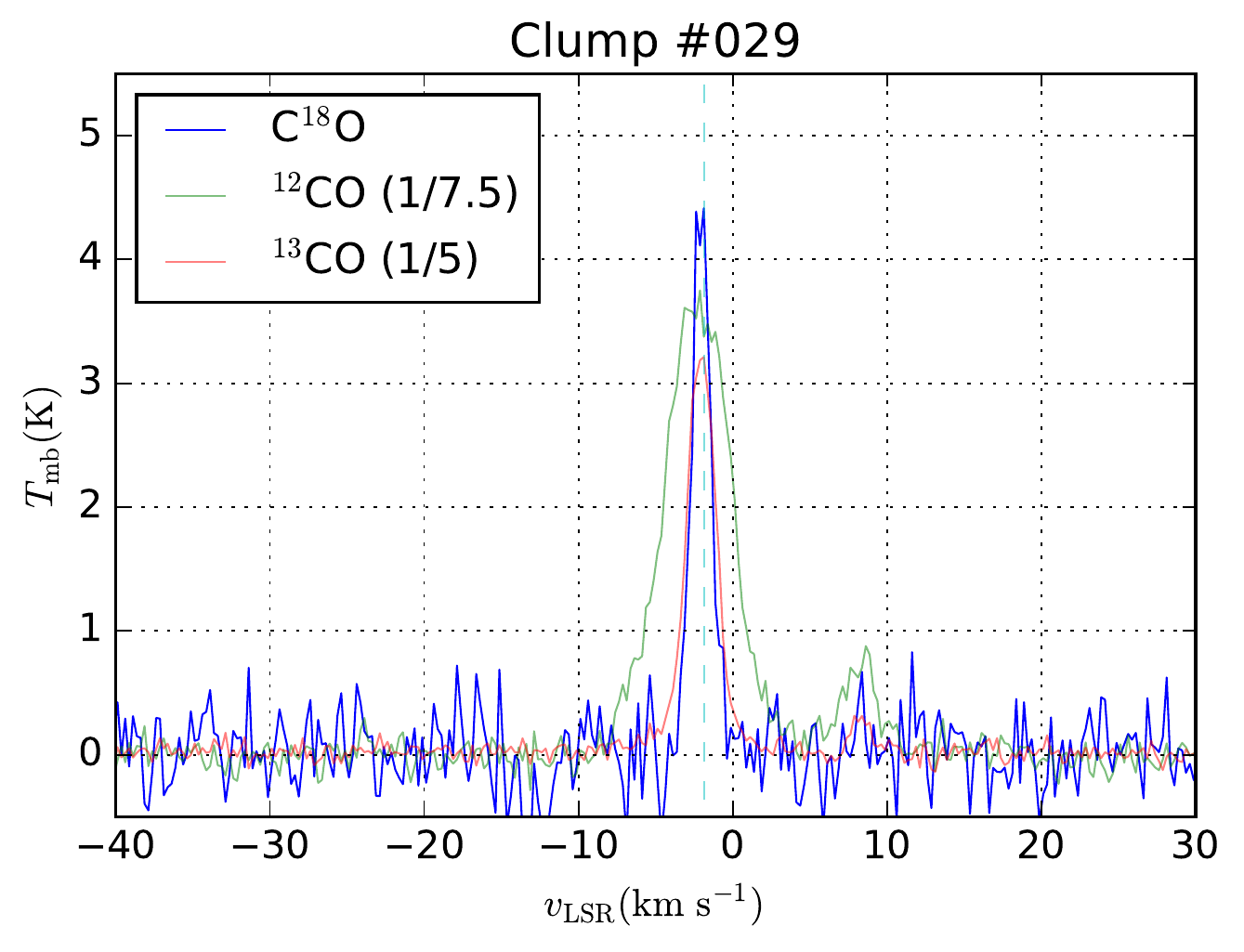}
\caption{Line profiles of the identified C$^{18}$O clumps. The spectrum of clump \#029 is shown as an example. The blue, green, and red solid lines represent C$^{18}$O, $^{12}$CO ($\times 1/7.5$), and $^{13}$CO ($\times 1/5$). The dashed cyan line shows the center velocity of the C$^{18}$O clumps. (The complete figure set (174 images) is available on the line.) \label{fig:lineprofile}}
\end{figure}

We show the line profiles of the C$^{18}$O clumps in Figure~\ref{fig:lineprofile}.
In the South region, we find that isolated clump \#3 shows a different peak velocity ($v_\mathrm{peak}=30~\mathrm{km~s^{-1}}$) from those of the other identified clumps.
The source position of \#3 corresponds to background star-forming region AFGL~2592 at $D=3.3$~kpc\citep{2012A&A...539A..79R}; thus, we exclude clump \#3 from the South objects.

For the statistical analysis presented in Section~\ref{sec4}, we did not use the eleven objects that were located at the edge of the map.
We also excluded the clumps in a patch around $(l,b)=(+78,-1)$, which included DR13S filaments \citep{2006A&A...458..855S}, in the statistical analysis to avoid a systematic bias in the physical properties owing to the difference in the map sensitivity.
We defined the border of the Cygnus~X North and South regions at a galactic longitude of $80.2 \deg$, with which the Cygnus OB2 association is associated.
Excluding the objects that were located in the DR13S regions and at the map edge, we used 68 and 65 C$^{18}$O clump samples as the North and South clumps, respectively. 

The star formation activities of the C$^{18}$O clumps were determined by existence of a protostar using the catalog obtained by \textit{Spitzer} \citep{2014AJ....148...11K}.
Thus, we found 98 clumps that were associated with one or more protostars and 35 starless clumps in the North and South regions that were not located at the map edge.

\section{Discussion}
\label{sec4}
In section~\ref{sec3}, we discussed the estimated physical properties of the C$^{18}$O clumps and the classification into North and South and star-forming and starless objects. 
Here, we discuss in detail the physical properties of the molecular cloud clumps identified by this survey.

\begin{deluxetable*}{lccccc}
\tabletypesize{\scriptsize}
\tablewidth{12cm}
\tablecaption{Averages and 1$\sigma$ deviations of physical properties of the C$^{18}$O clumps.\label{table2}}
\tablehead{&All&North&South&Star-forming&Starless}
\startdata
Samples&133&68~(51.1 \%)&65~(48.9 \%)&98~(73.7 \%)&35~(26.3 \%)\\
$R_\mathrm{cl}$(pc)&0.49 $\pm$0.17 &0.49 $\pm$0.17 &0.49 $\pm$0.17 &0.52 $\pm$0.17 &0.42 $\pm$0.13 \\
$dV_\mathrm{cl}$($\mathrm{km\ s^{-1}}$)&0.75 $\pm$0.39 &0.79 $\pm$0.42 &0.71 $\pm$0.37 &0.83 $\pm$0.40 &0.54 $\pm$0.29 \\
$M_\mathrm{LTE}$ ($10^2\ \mathrm{M_\odot}$)&1.9 $^{+3.0 }_{-1.1 }$&2.2 $^{+3.6 }_{-1.4 }$&1.5 $^{+2.3 }_{-0.9 }$&2.3 $^{+3.6 }_{-1.4 }$&1.0 $^{+1.1 }_{-0.5 }$\\
$n_\mathrm{H_2}$($10^3\mathrm{cm^{-3}}$)&6.6 $^{+6.1 }_{-3.2 }$&7.9 $^{+7.8 }_{-3.9 }$&5.5 $^{+4.2 }_{-2.4 }$&7.1 $^{+6.7 }_{-3.4 }$&5.4 $^{+4.3 }_{-2.4 }$\\
$\alpha_\mathrm{vir}$&0.30 $\pm$0.24 &0.28 $\pm$0.21 &0.33 $\pm$0.27 &0.32 $\pm$0.41 &0.27 $\pm$0.29 \\
\enddata
\end{deluxetable*}

\subsection{Difference in the physical properties of the North and South regions}
Although both the Cygnus~X North and South regions are known as large reservoirs of molecular gases ($\sim 10^6~M_\sun$), the North region shows an extremely filamentary $^{12/13}$CO structure compared to the South region \citep{2006A&A...458..855S, 2011A&A...529A...1S} and contains numerous active star-forming regions represented by DR21 and W75N.
The above seem to reflect a difference in the star formation activity and evolution stages of each molecular cloud complex, and therefore, a difference in the statistical properties of the C$^{18}$O clumps of the North and South regions is suggested.

\begin{figure*}
\epsscale{1}
\plotone{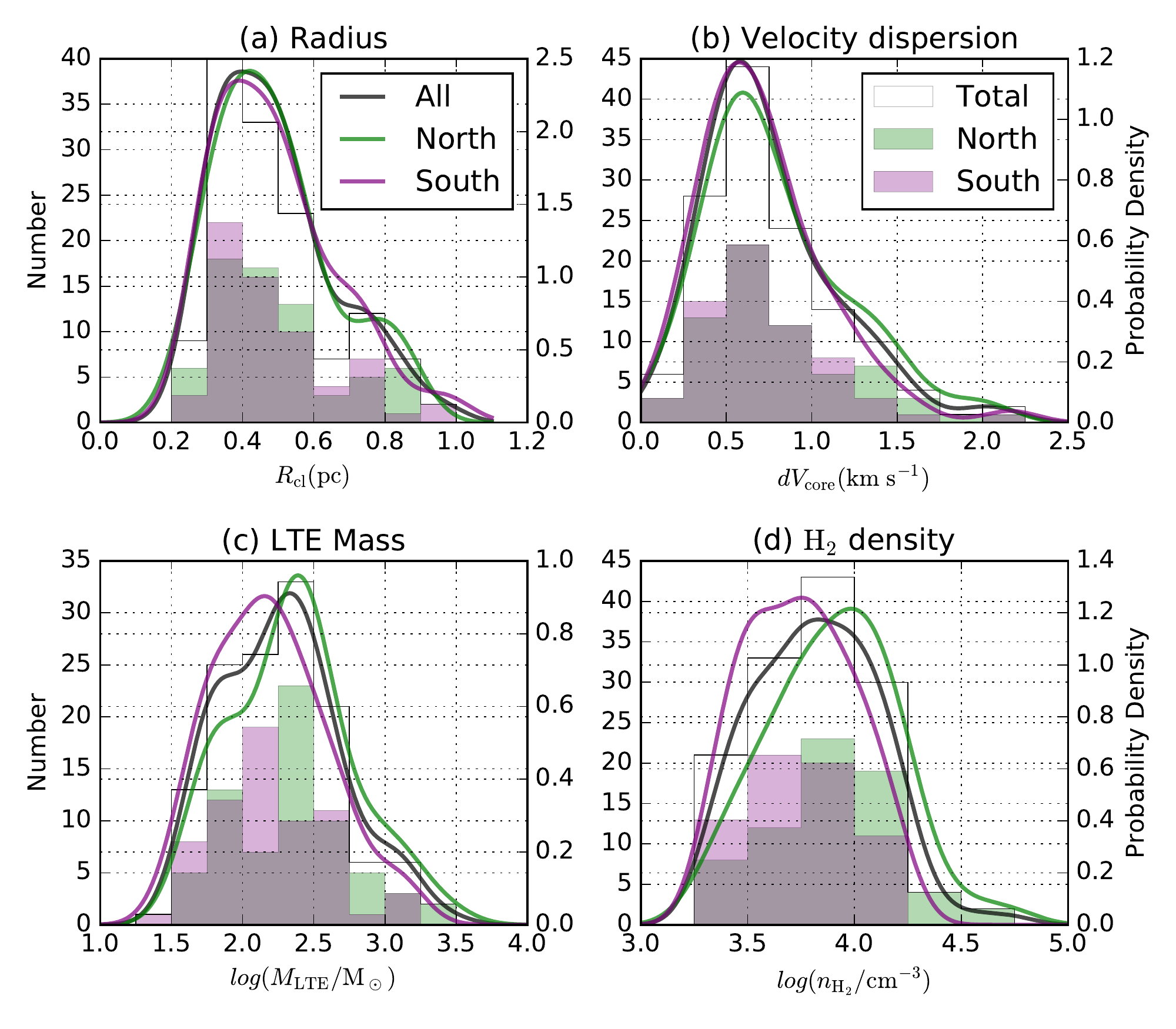}
\caption{Probability densities estimated by the kernel density estimation with Scott's rule of the bandwidth and histograms of the (a) radius, (b) velocity dispersion, (c) LTE mass, and (d) $\mathrm{H_2}$ density of the C$^{18}$O clumps located in the North and South regions. The green, purple, and black solid lines show the probability densities of the North, South, and all the (North and South) clumps, respectively. The green, purple, and white rectangular regions show the histograms of the North, South, and all the clumps, respectively.\label{fig:histNS}}
\end{figure*}

Figure~\ref{fig:histNS} shows the probability densities and histograms of the physical properties of the identified C$^{18}$O clumps of the North and South regions.
The mean and standard deviation of the physical properties classified by the regions are listed in Table~\ref{table2}.
The distributions of the radius and velocity dispersion are very similar for the North and South clumps. 
By contrast, the C$^{18}$O clumps in the North region show a slightly larger LTE mass and higher $\mathrm{H_2}$ density than those in the South region.
The $p$-values of Welch's $t$-test for the radii, velocity dispersions, LTE masses, and $\mathrm{H_2}$ densities of the North and South cores are 0.959, 0.241, 0.070, and 0.002, respectively.
This does not support that the average values of the radii and velocity dispersions of the North and South clumps are significantly different.
Contrastingly, the average $\mathrm{H_2}$ density of the C$^{18}$O clumps of the North region is significantly higher than that of the C$^{18}$O clumps of the South region, with a significance level of 5\%.
Thus, we can expect that the statistical difference in the $\mathrm{H_2}$ density reflects that of the clump evolution stages, and therefore, the difference in the star formation activities of the North and South regions, as suggested by \citet{2006A&A...458..855S} and \citet{2018ApJS..235....9Y}.

\subsection{Difference in the physical properties by the presence of star formation activity}
\label{sec4.2}
We also investigated the quantitative difference in the physical properties of star-forming and starless clumps, which seemed to reflect the evolution sequences of the C$^{18}$O clumps.

\begin{figure*}
\epsscale{1}
\plotone{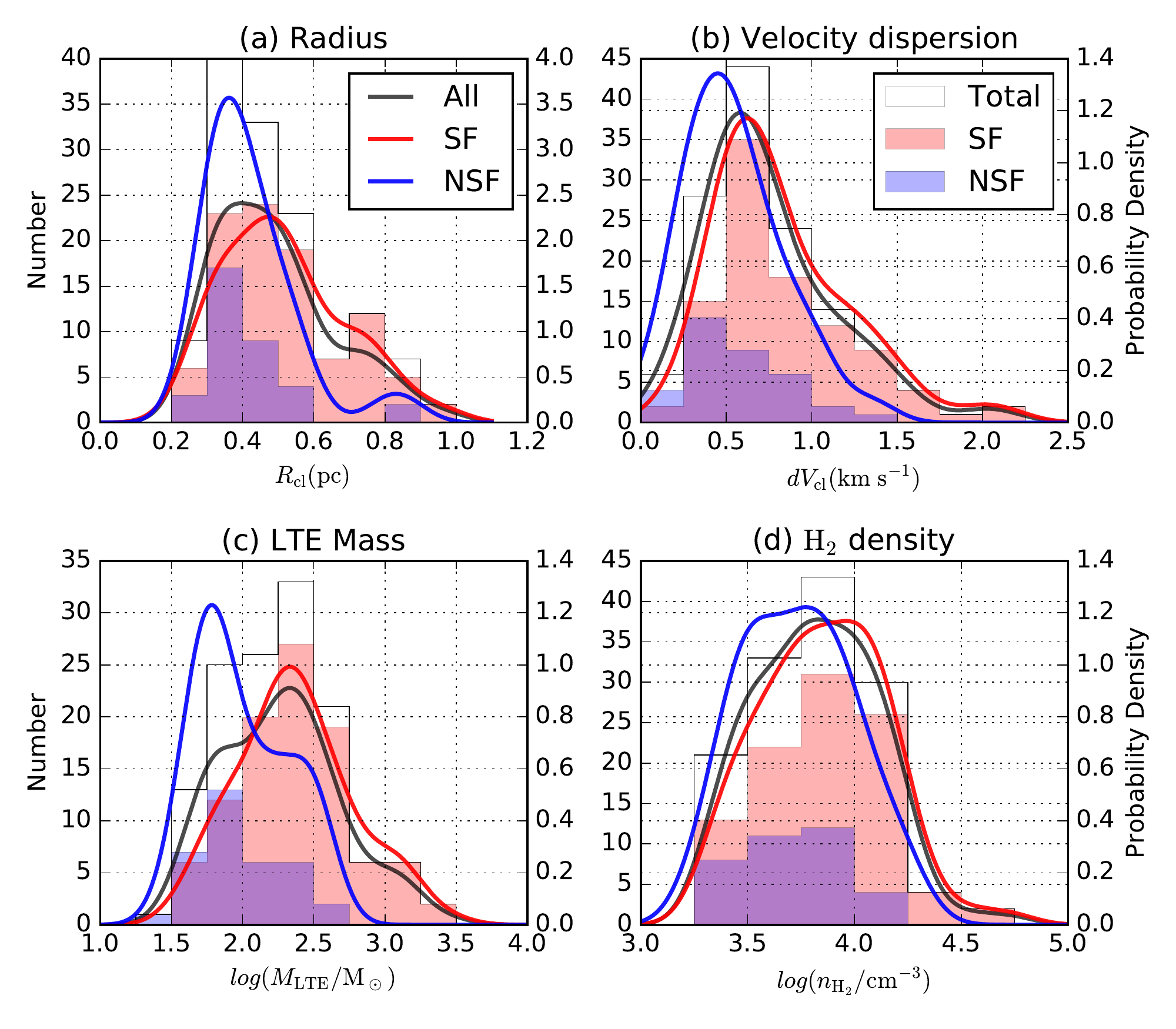}
\caption{Probability densities estimated by the kernel density estimation with Scott's rule of bandwidth and histograms of the (a) radius, (b) velocity dispersion, (c) LTE mass, and (d) $\mathrm{H_2}$ density of the C$^{18}$O clumps classified by the existence of a protostar to determine the presence of star-formation activity. The red, blue, and black solid lines show the probability densities of the star-forming, starless, and all the (both North and South) clumps, respectively. The red, blue, and white rectangular regions show the histograms of the star-forming, starless, and all the clumps, respectively.\label{fig:histSF}}
\end{figure*}

Figure~\ref{fig:histSF} shows the probability densities and histograms of the physical properties of the identified C$^{18}$O clumps of the star-forming and starless clumps.
In Table~\ref{table2}, we also list the average and standard deviation classified by the star formation activity of the clumps.
The star-forming clumps show a larger radius, velocity dispersion, LTE mass, and $\mathrm{H_2}$ density than the starless clumps.
In fact, the significant difference in the mean values of these properties is strongly supported by Welch's $t$-test, whose $p$-values are $<$0.01, $<$0.01, $<$0.01, and 0.01, respectively.

In nearby ($\leq200$~pc) star-forming regions, \citet{2002A&A...385..909T} reported a similar tendency of the C$^{18}$O core radii, velocity dispersions, LTE masses, $\mathrm{H_2}$ densities, and $\mathrm{H_2}$ column densities between the starless, star-forming, and cluster-forming cores.
This tendency is naturally expected to arise from difference of gas accretion time scale between starless and star-forming cores.
Our result also supports that the general trends of the core properties, which evolve with star formation, are applicable to the C$^{18}$O clumps in an extremely active high-mass star-forming region.

\subsection{Virial ratio}
Here, we discuss the virial ratio, which is an important indicator to determine whether stars will form cores/clumps, in comparison with previous C$^{18}$O studies.
As references of the previous C$^{18}$O surveys, we use the results of nearby ($D\leq200$~pc) molecular clouds, including the low-mass star-forming regions of Taurus, Ophiuchus, Lupus, Lynds~1333, Corona Australis, Southern Coalsack, and the Pipe nebula, observed by NANTEN\citep[][]{2002A&A...385..909T}.
We also refer to the C$^{18}$O cores properties observed by the NRO~45m telescope in Orion~A \citep[$D=440$~pc,][]{2015ApJS..217....7S, 2007PASJ...59..897H}, the nearest high-mass star-forming GMC, which has about 1/10 of the total molecular gas mass of the Cygnus~X GMC complex \citep{2018ARA&A..56...41M}.
We also consider the Sharpless 2-140  (S~140) H{\small II} region, a compact high-mass star-forming region located at the edge of the Lynds~1204 molecular cloud \citep[$D=760$~pc,][]{2011ApJ...732..101I,2008PASJ...60..961H}.
The above is done to compare with smaller and less active high-mass star-forming regions than Cygnus~X.
We followed the assumption of a uniform and spherical core structure that had no support of rotation, magnetic field, and external pressure, which is assumed in the previous studies.
By considering the most compact clumps ($n_\mathrm{vox}=16$) with relatively high excitation temperature of 36~K, corresponds to the $2\sigma$ value of all samples, the detection limit of the LTE mass is estimated to be $\sim 55~M_\sun$ with $3\sigma$ intensity detection, which covers the 95th percentile of the identified clumps.
Therefore, we defined 95th percentile as induces of relative sensitivity limit for the Cygnus X, S140, Orion A, and nearby low-mass samples.
Thus, the relative sensitivity limits of the LTE masses in the Cygnus~X samples are $\sim$10 and 20 times worse than S~140, and Orion~A/nearby studies.

We also consider the systematic bias of $\alpha_\mathrm{vir}$ caused by different spatial and velocity resolutions of data set.
From the definition of the LTE and virial masses, we can assume that $\alpha_\mathrm{vir}$ is proportional to spatial resolution and inversely proportional to velocity resolution.
The spatial and velocity resolutions of the Cygnus~X clumps were 3 and 2.5 times worse than previous core studies.
Thus, systematic bias of the virial ratio estimate can be estimated to be a factor of $\sim$1.2, and would not affect our discussion.

\begin{figure}
\epsscale{1}
\plotone{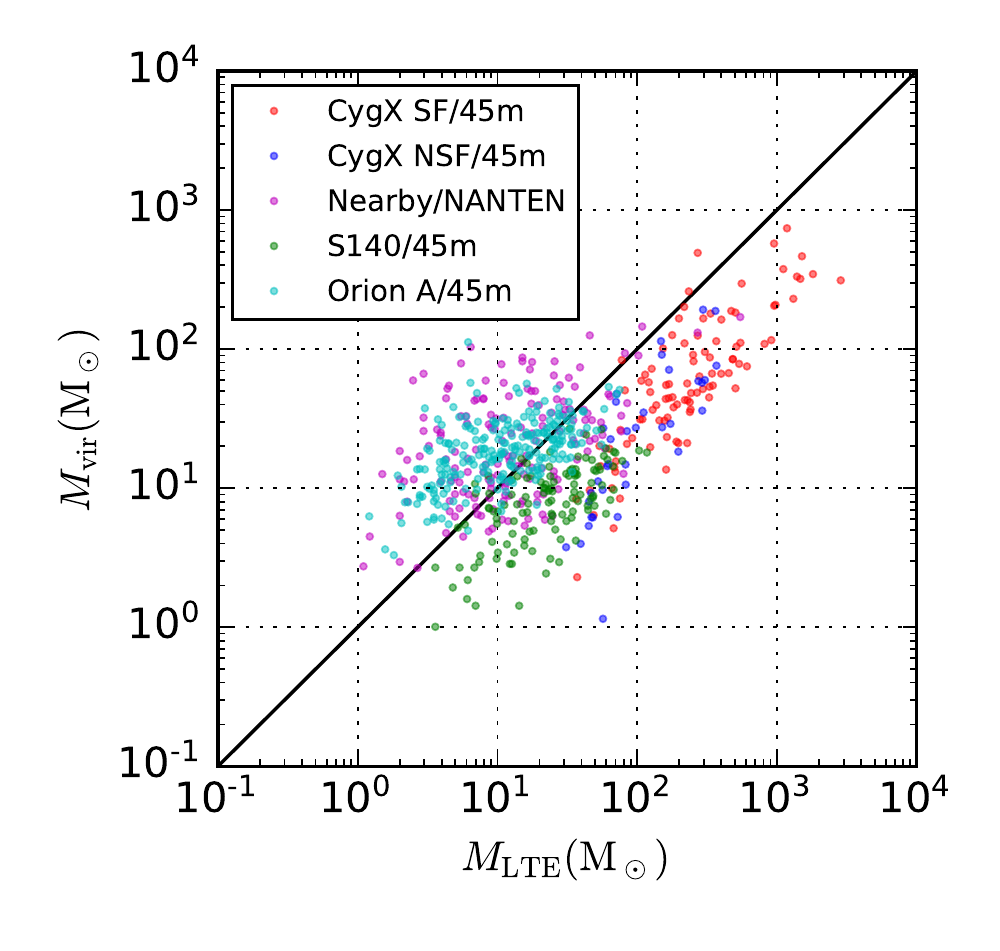}
\caption{Relations between the LTE mass and virial mass. The red, blue, magenta, green, and cyan points show the Cygnus~X star-forming, starless, NANTEN nearby molecular clouds, S~140, and Orion~A C$^{18}$O clumps/cores, respectively.
The black solid line shows the relation of $\alpha_\mathrm{vir}=1$. \label{fig:LTEvsVIR}}
\end{figure}

The relation between the LTE and virial masses is shown in Figure~\ref{fig:LTEvsVIR}.
While the NANTEN and Orion~A C$^{18}$O cores are located at $\alpha_\mathrm{vir}\ga 1$, most of the C$^{18}$O core/clumps in Cygnus~X and S~140 show a virial ratio of $\alpha_\mathrm{vir}<1$.
The average and standard deviation values of the virial ratio of the star-forming and starless clumps in Cygnus~X are $0.32\pm0.26$, $0.27\pm0.19$ respectively, and the difference is not significant with Welch's $t$-test ($p=0.32$).
The average value of both the star-forming and starless cores is $0.30\pm0.24$, which is consistent with the virial ratio of S~140 ($\alpha_\mathrm{vir}=0.35\pm0.23$) and, therefore, supports that these C$^{18}$O clumps are gravitationally bound.
The observing region of S~140 is only $20\arcmin \times 18\arcmin$, and it will be very biased to the center of the active star-forming region.
This result suggests that the C$^{18}$O clumps in Cygnus~X have very similar properties at the center of the high mass star-forming region. Thus, the C$^{18}$O clumps in Cygnus~X traces a dense molecular gas clump that directly connects to an extremely active future and current star formation activity.
In addition, the virial ratios in Cygnus~X and S~140 are smaller than those in the nearby molecular clouds ($\alpha_\mathrm{vir}=2.8\pm 3.6$) and Orion~A ($\alpha_\mathrm{vir}=2.4\pm2.2$).
This might reflect the difference in the star formation mode: low-mass single star or high-mass cluster formation.

Another important feature is that the distribution of the virial or LTE mass of the C$^{18}$O clumps in Cygnus~X is dispersed widely compared to those in the previous studies.
In fact, the mean and standard deviation of LTE masses in the previous studies were $22^{+25}_{-12}$, $11^{+14}_{-6}$, and $12^{+23}_{-8}~\mathrm{M_\sun}$ in S~140, Orion~A, and nearby star-forming regions, respectively. 
These values are much smaller than those in the Cygnus X ($190^{+300}_{-110}~\mathrm{M_\sun}$) by more than one order of magnitude.
This result is consistent with our estimate of the relative sensitivity limits, which would be attribute to the lower spatial resolution, spectral resolution, and image sensitivity than the previous studies.

While the typical LTE and virial masses are larger than the previous studies, the fact that the clumps in Cygnus~X show $\alpha_\mathrm{vir}< 1$ supports that these clumps are gravitationally bound objects, which are directly related to star formation.
In particular, some of the massive clumps are assumed to be the formation sites of high-mass stars and stellar clusters.
Thus, this feature could be related to the extremely active star formation in Cygnus~X.

\begin{deluxetable}{llcc}
\tabletypesize{\scriptsize}
\tablewidth{12cm}
\tablecaption{Fitting results of the CMF parameters.\label{table3} \textcolor{blue}{[This table was updated.]}}
\tablehead{Sample & Fitting mass range & $\alpha$ & $a~(\times 10^3)$}
\startdata
All & $1.75<\log(M_\mathrm{LTE}/M_\odot)<2.15$ & -1.39 $\pm$ 0.04 & $0.57\pm 0.10$ \\
All & $2.15<\log(M_\mathrm{LTE}/M_\odot)$ & -2.07 $\pm$ 0.04 & $17.4\pm 3.6$ \\
Star-forming & $1.75<\log(M_\mathrm{LTE}/M_\odot)<2.15$ & -1.30 $\pm$ 0.04 & $0.31\pm 0.05$ \\
Star-forming & $2.15<\log(M_\mathrm{LTE}/M_\odot)$ & -2.00 $\pm$ 0.04 & $10.6 \pm 2.2$ \\
Starless & $1.75<\log(M_\mathrm{LTE}/M_\odot)$ & -1.94 $\pm$ 0.06 & $1.19 \pm 0.37$ \\
\enddata
\end{deluxetable}

\subsection{Clump mass function}
We also examined the core/clump mass function (CMF) in Cygnus~X to reveal the detailed mass properties and relation between the IMF and galactic field stars.
Based on the definition of \citet{2014prpl.conf...53O}, the IMF and CMF are defined as 
\begin{equation}
\frac{dN}{dM} \propto {M}^{\alpha},
\end{equation}
where $N$ is the number of stars or cores/clumps, $M$ is the mass of the stars or cores/clumps, and $\alpha$ is the spectral index of the IMF or CMF.
As an integral form of the CMF for the case of $\alpha<-1$, we define a cumulative number,
\begin{equation}
N(>M_\mathrm{LTE}) \equiv \int^{\infty}_{M_\mathrm{LTE}}\frac{dN}{dM}dM= a {M_\mathrm{LTE}}^{\alpha+1},
\end{equation}
where $a$ is the factor of proportionality.

The observational studies of dense dust core surveys using (sub)millimeter dust continuum and dust extinction \citep[e.g.,][]{1998A&A...336..150M,2006ApJ...638..293E,2007MNRAS.374.1413N,2007A&A...462L..17A} have revealed that multiple spectral index components of the CMF are similar to the $\alpha=-1.3\pm0.5$ ($0.08<M_*/M_\sun<0.5$, $M_*$ is a stellar mass) and $\alpha=-2.3\pm0.3$ ($0.5<M_*/M_\sun<1$) components of the Kroupa IMF\citep{2001MNRAS.322..231K}.
For a C$^{18}$O core observation, \citet{2002A&A...385..909T} has also reported multiple spectral index components of the CMF ($\alpha=1.25$ and $2.5$) corresponding to the Kroupa IMF components at $0.08<M_*/M_\sun<0.5$ and $0.5<M_*/M_\sun<1$ ranges toward nearby low-mass star-forming regions.
Our observation in Cygnus~X provides large samples of C$^{18}$O clumps in an extremely active cluster-forming region.
\citet{2003ARA&A..41...57L} suggest that cluster formation activity in GMCs is the dominant (70\%--90\%) supplier of field low-mass stars in the galactic disk.
Thus, it is important to investigate the relationship between the IMF of the galactic field stars and CMF obtained by our C$^{18}$O clump samples, which are more massive and larger than those in the previous C$^{18}$O studies. 

\begin{figure}
\epsscale{1}
\plotone{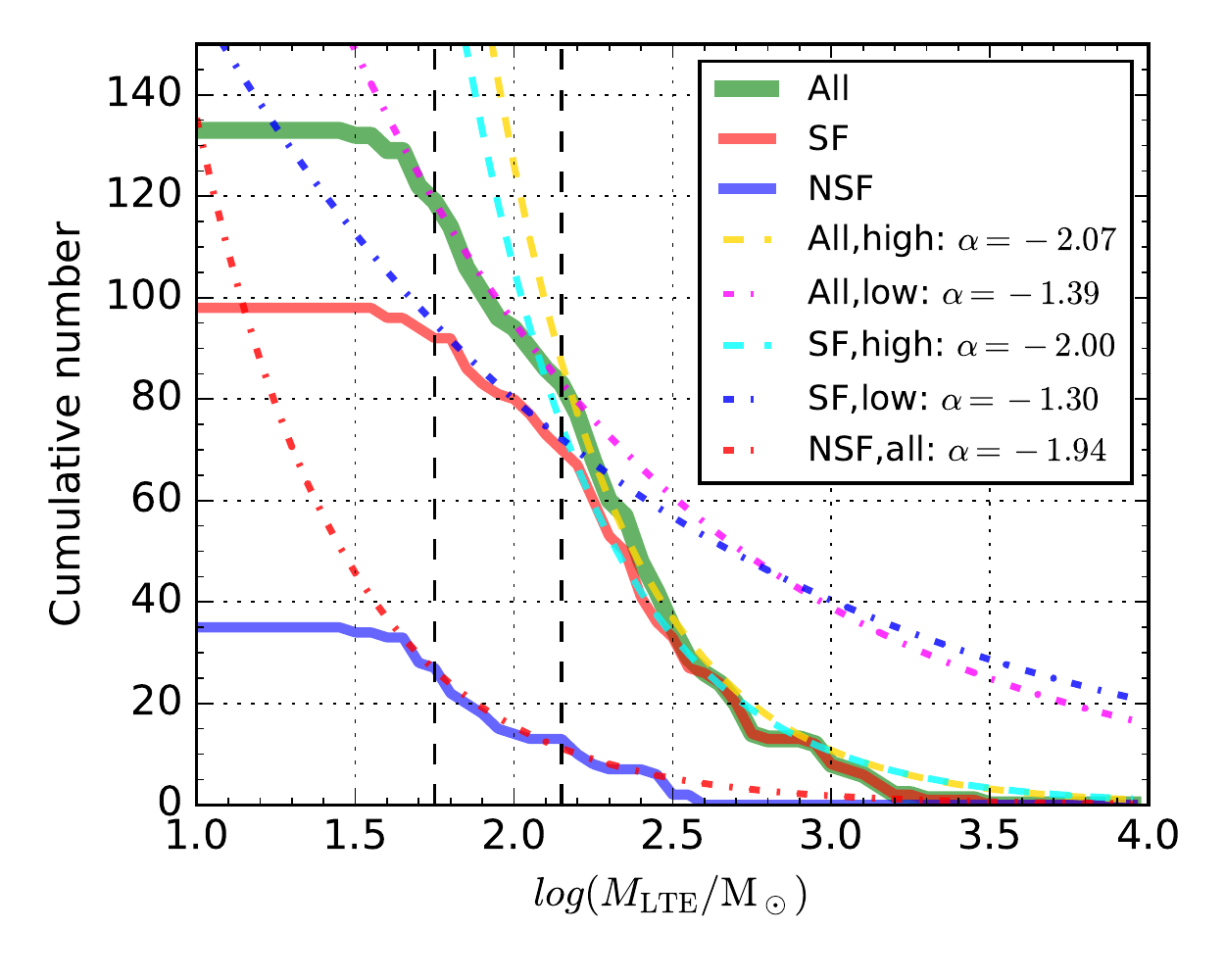}
\caption{Cumulative numbers of $M_\mathrm{LTE}$ of the total, star-forming (SF), and starless (NSF) C$^{18}$O clumps shown in green, red, and blue solid line, respectively. The best-fit CMFs for the high- and low-mass parts are shown in dashed and dot-dashed lines, respectively. The vertical dashed lines show the detection limit, $\log(M_\mathrm{LTE}/M_\odot)=1.75$, and the boundary mass changing the spectral induces of the CMFs, $\log(M_\mathrm{LTE}/M_\odot)=2.15$. The fitting parameters are listed in Table~\ref{table3}. \label{fig:massf}}
\end{figure}

Figure~\ref{fig:massf} shows the cumulative number count of the C$^{18}$O clumps and suggests that the spectral index of the CMF changes around $\log(M_\mathrm{LTE}/M_\odot)=2.15$.
The fitting parameters are shown in Figure~\ref{table3}.
Considering the detection limit of the C$^{18}$O clumps, $\log(M_\mathrm{LTE}/M_\odot)=1.75$, we fit with two mass functions with ranges of $1.75<\log(M_\mathrm{LTE}/M_\odot)<2.15$ and $2.15<\log(M_\mathrm{LTE}/M_\odot)$.
From the least-$\chi^2$ fittings, which are shown in Figure~\ref{fig:massf}, the cumulative number count of the C$^{18}$O clumps fits well in each mass range.
We obtained $\alpha=-1.39\pm0.04$ for $1.75<\log(M_\mathrm{LTE}/M_\odot)<2.15$, and $\alpha=-2.07\pm0.04$ for $2.15<\log(M_\mathrm{LTE}/M_\odot)$.
The errors (1$\sigma$) were estimated by Monte Carlo simulation by considering the random errors of the estimated LTE masses of the C$^{18}$O clumps.
These spectral indices are consistent with the $\alpha=-1.3$ and $-2.3$ components of the Kroupa IMF.
This confirms the similarity of the IMF in the galactic field stars and clump-scale CMF in a high-mass star-forming region.

We also investigate the difference in the spectral indices of the star-forming and starless clumps. 
The mass function of the star-forming clumps well fits the two components of the spectral indices: $\alpha=-1.30\pm0.04$ for $1.75<\log(M_\mathrm{LTE}/M_\odot)<2.15$ and $\alpha=-2.00\pm0.04$ for $2.15<\log(M_\mathrm{LTE}/M_\odot)$, which are also consistent with the IMF.
However, for the starless clumps, we can fit the CMF with a single spectral index of $\alpha=-1.94\pm0.06$ at the mass range of $1.75<\log(M_\mathrm{LTE}/M_\odot)$, and the index of the starless clumps is consistent with the spectral index of a star-forming clump at a high mass range.
Thus, we can assume that the starless clumps will evolve into star-forming clumps with further gas mass accretion.
This is also supported by the fact that $\sim2$ times lowers the average mass of the starless clumps more than that of the star-forming ones, as can be seen from Table~\ref{table2}.

\subsection{Star formation efficiency of the C$^{18}$O clumps}
We can estimate the molecular gas mass fraction that contribute to the stellar mass in a clump, which is called the star formation efficiency (SFE), from the boundary gas mass that changes the spectral index.
Here, we assume that the clumps with a boundary mass of $\log(M_\mathrm{LTE})=2.15$ (i.e., $M_\mathrm{LTE}\simeq 140~M_\odot$) evolve into a single star that has a boundary mass of the Kroupa IMF of 0.5~$M_\sun$ or into a cluster that has a maximal stellar mass of 0.5~$M_\sun$.
In case cluster formation, using the relation between maximal stellar mass $M_\mathrm{*,max}$ and cluster mass $M_\mathrm{cluster}$: $M_\mathrm{*,max}=0.39 M_\mathrm{cluster}^{2/3}$, assuming the hierarchical cluster formation model \citep{2003MNRAS.343..413B, 2004MNRAS.349..735B, 2006MNRAS.365.1333W,2010MNRAS.401..275W}, the total cluster stellar mass of the cluster is expected to be $1.5~M_\sun$.
Thus, the SFE of the typical C$^{18}$O clumps is expected to be 0.3--1\%. This is very unlikely because the SFE is excessively lower than that estimated for the low-mass star-forming regions observed by NANTEN from the comparison of the CMF and IMF, with the assumption of a single star formation in the C$^{18}$O cores \citep[$\sim$10\%,][]{2002A&A...385..909T}.

It is known that some studies of massive clumps also reveal a high SFE \citep[$\sim$10\%, e.g.,][]{2003ARA&A..41...57L} by comparing the gas amount with the stellar content in GMCs.
Assuming an SFE of 10\%, the $\sim$10 C$^{18}$O clumps that have gas masses of $\ga 10^3~M_\sun$ will evolve into open clusters having a total stellar mass of $\ga 100~M_\sun$ and containing one or more high-mass stars ($> 8~M_\sun$).
This scenario is consistent with a high-resolution interferometry study of massive dense cores in Cygnus~X North \citep{2010A&A...524A..18B}, which revealed numerous fragmentary structures inside massive dense cores.

The discrepancy between the SFEs of the NANTEN C$^{18}$O cores and our samples could be explained in terms of the physical spatial resolution of our dataset of Cygnus~X ($\sim$0.3~pc) being worse than those of the NANTEN observations ($\sim$0.1~pc). This is because the identified C$^{18}$O clumps in Cygnus~X are larger than in the NANTEN study, and therefore, the mass of these clumps is higher than of those in the latter study.
Thus, we can also expect that most of the C$^{18}$O clumps in Cygnus~X have an internal structure, and our predicted SFE using the relation of the IMF and CMF might be underestimated.
Further high-resolution, high-sensitivity, and wide-field survey of C$^{18}$O and other dense gas tracers toward high-mass star-forming regions is important to understand the complete mechanism of star formation across a GMC.

\section{Summary}
\label{sec5}
We investigated the physical properties of the C$^{18}$O clumps identified in a multi-line CO ($J=$1--0) survey toward the Cygnus~X regions using the Nobeyama 45-m radio telescope.
The main results are summarized below.
\begin{enumerate}
\item{We identified 174 C$^{18}$O clumps in total. Ninety eight out of the 133 objects, except for the objects that were located at the map edge or in the DR13S region, were accompanied by one or more protostars.}
\item{The C$^{18}$O clump properties showed clump radii of 0.2--1~pc, velocity dispersions of $<2.2 \mathrm{km\ s^{-1}}$, LTE masses of 30--3000~$M_\sun$, and H$_2$ densities of (2--55)~$\times 10^3\ \mathrm{cm^{-3}}$.}
\item{We detected statistical differences in the physical properties of the clumps of the North and South regions in terms of the H$_2$ density. This was consistent with difference in the actual star formation activities of these regions and suggested to be caused by the difference in the evolution stages in the North and South regions.}
\item{The statistical differences in the physical properties of the star-forming and starless clumps were confirmed to be significant. The larger radius and velocity dispersion and higher LTE mass and H$_2$ density in the star-forming clumps compared to those in the starless ones reflected the difference in the clump evolution stages.}
\item{The average value of the virial ratio was $0.30\pm0.24$. This supported that the C$^{18}$O clumps in Cygnus~X were gravitationally bound and served as formation sites of a star or stellar cluster. In addition to Cygnus~X, an active cluster-forming region, S~140, also reported a lower virial ratio than the nearby low-mass star-forming molecular clouds and Orion~A GMC. This tendency seemed to be characterized by the difference in the star formation mode in these observing regions.}
\item{We confirmed two spectral index components of the clump-scale CMF, $\alpha=-1.39\pm0.04$ ($1.75<\log(M_\mathrm{LTE}/M_\odot)<2.15$) and $\alpha=-2.07\pm0.04$ ($2.15<\log(M_\mathrm{LTE}/M_\odot)$), which were consistent with the $\alpha=-1.3$ ($0.08<M_*/M_\sun<0.5$, $M_*$) and $\alpha=-2.3$ ($0.5<M_*/M_\sun<1$) components of the IMF of the galactic field stars, respectively.}
\item{The mass function spectral index of the star-forming clumps, $\alpha=-2.00\pm0.04$, at $2.15<\log(M_\mathrm{LTE}/M_\odot)$ was consistent with that of the starless clumps, $\alpha=-1.94\pm0.06$ at $1.75<\log(M_\mathrm{LTE}/M_\odot)$, suggesting that the starless clumps would evolve into star-forming clumps with further gas mass accretion.}
\item{By comparing the boundary masses of the CMF and IMF, the SFE of the C$^{18}$O clumps was estimated to be 0.3--1\%, which was excessively lower than that reported in previous studies ($\sim$10\%) and very unlikely. 
Assuming an likely SFE of 10\%, about ten C$^{18}$O clumps that had a gas mass of $>10^3~M_\sun$ were expected to evolve into open clusters containing one or more high-mass stars.}
\end{enumerate}

\acknowledgments
K.~T. would like to thank the University of Virginia for providing the funds for her postdoctoral fellowship in the VICO research program.
The Nobeyama 45-m radio telescope is operated by Nobeyama Radio Observatory, a branch of the National Astronomical Observatory of Japan.
Data analysis was in part carried out on the open-use data analysis computer system at the Astronomy Data Center, ADC, of the National Astronomical Observatory of Japan.
This research made use of Astropy, a community-developed core Python package for astronomy \citep[\url{http://www.astropy.org/},][]{2013A&A...558A..33A}, and \texttt{astrodendro}, a Python package to compute dendrograms of astronomical data (\url{http://www.dendrograms.org/}).
This work was supported by JSPS KAKENHI Grant Numbers JP17H06740, JP18K13580, JP18K13582, and JP18K13595.

%

\vspace{5mm}
\facilities{No:45m, Spitzer}


\software{Astropy\citep{2013A&A...558A..33A}, NumPy\citep{walt2011numpy}, SciPy\citep{scipy2001}, Matplotlib\citep{Hunter:2007}, astrodendro}

\bibliographystyle{yahapj}



\end{document}